%% file: iclr2026_conference.tex
\documentclass{article}
\PassOptionsToPackage{x11names}{xcolor}
\usepackage{iclr2026_conference,times}
\usepackage[normalem]{ulem}

\input{math_commands.tex}

\usepackage{hyperref}
\usepackage{url}

\usepackage[utf8]{inputenc}
\usepackage[T1]{fontenc}
\usepackage{booktabs}
\usepackage{amsfonts}
\usepackage{nicefrac}
\usepackage{microtype}
\usepackage{xspace}

\usepackage{subcaption}
\usepackage{caption}

\usepackage[pdftex]{graphicx}
\graphicspath{{images/}}
\usepackage{rotating}

\usepackage{siunitx}
\usepackage{multirow}
\usepackage{makecell}
\usepackage{array}
\usepackage{xcolor}
\usepackage{colortbl}
\definecolor{skyhex}{HTML}{4E95D9}
\definecolor{orangebox}{HTML}{FDE7D0}
\definecolor{greenbox}{HTML}{D9F2D0}
\definecolor{bluebox}{HTML}{D4F0EA}
\definecolor{highlightyellow}{HTML}{FFF59D}

\definecolor{darkblue}{rgb}{0, 0, 0.5}
\hypersetup{colorlinks=true, citecolor=darkblue, linkcolor=darkblue, urlcolor=darkblue}

\usepackage[most]{tcolorbox}
\usepackage{listings}
\usepackage[newfloat,frozencache,cachedir=.]{minted}
\newenvironment{code}{\captionsetup{type=listing}}{}
\SetupFloatingEnvironment{listing}{name=Listing}
\usepackage{titletoc}
\usepackage{fontawesome}
\newtcolorbox{promptbox}[1][]{%
  colback=blue!10,
  colframe=blue!30!black,
  fonttitle=\bfseries,
  title=#1,
  sharp corners,
  breakable,
  fontupper=\small
}

\makeatletter
\setminted{
  linenos,
  breaklines,
  numbersep=5pt,
  xleftmargin=\dimexpr\@totalleftmargin\relax
}
\makeatother

\usepackage{verbatim}
\usepackage{float}

\title{\cresearcher{}: Deep Research Agent for\\ Large Systems Code and Commit History}

\author{%
    Ramneet~Singh\thanks{Equal contribution} \quad Sathvik~Joel\footnotemark[1] \quad Abhav~Mehrotra \quad Nalin~Wadhwa \\ \textbf{Ramakrishna~B~Bairi \quad Aditya~Kanade \quad Nagarajan~Natarajan} \\
    Microsoft Research \\
    \texttt{\{ramneet2001,ksjoe30,abhavm1,nalin.wadhwa02\}@gmail.com} \\
    \texttt{\{ram.bairi,kanadeaditya,nagarajan.natarajan\}@microsoft.com}
}

\newcommand{\cresearcher}{\textsl{Code Researcher}}
\newcommand{\kbenchsyz}{kBenchSyz}
\newcommand{\kgym}{\textsl{kGym}}
\newcommand{\kbuilder}{\textsl{kBuilder}}
\newcommand{\kreproducer}{\textsl{kReproducer}}
\newcommand{\kscheduler}{\textsl{kScheduler}}
\newcommand{\kdashboard}{\textsl{kDashboard}}
\newcommand{\kmq}{\textsl{kmq}}

\newcommand{\analysisphase}{\textsc{Analysis}\xspace}
\newcommand{\synthesisphase}{\textsc{Synthesis}\xspace}
\newcommand{\validationphase}{\textsc{Validation}\xspace}
\newcommand{\maxcalls}{max calls\xspace}
\newcommand{\ffmpeg}{FFmpeg\xspace}

\iclrfinalcopy
\begin{document}

\maketitle

\vspace{-0.7cm}

\begin{figure}[h!]
    \centering
    \includegraphics[width=\linewidth]{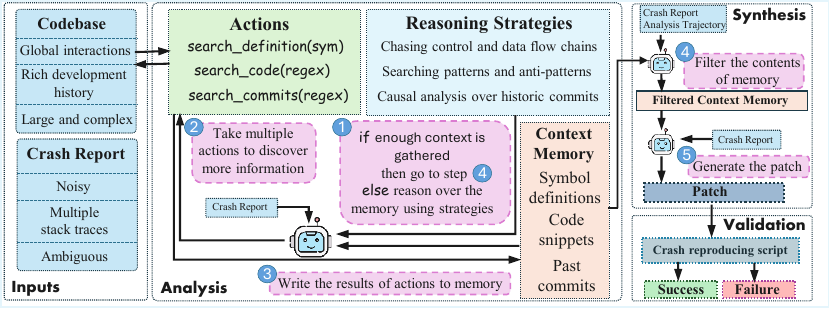}
    \caption{\cresearcher{} conducts deep research over code in three phases: (1) Starting with the codebase and crash report as input, the \analysisphase{} phase performs multi-step reasoning about semantics, patterns, and commit history of code. It gathers context in a memory. (2) The \synthesisphase{} phase filters the contents of the memory to keep relevant context and generates a patch. (3) The \validationphase{} phase uses external tools to validate the patch.}
    \label{fig:design}
\end{figure}

\begin{abstract}
    Large Language Model (LLM)-based coding agents have shown promising results on coding benchmarks, but their effectiveness on systems code remains underexplored. Due to the size and complexities of systems code, making changes to a systems codebase requires \emph{researching} about many pieces of context, derived from the large codebase and its massive commit history, \emph{before} making changes. Inspired by the recent progress on deep research agents, we design the first deep research agent for code, called \cresearcher{}, and apply it to the problem of generating patches to mitigate crashes reported in systems code. \cresearcher{} performs multi-step reasoning about semantics, patterns, and commit history of code to retrieve all relevant context from the codebase and its commit history. 
    We evaluate \cresearcher{} on \kbenchsyz{}~\citep{kgym}, a benchmark of Linux kernel crashes, and show that it significantly outperforms strong baselines, achieving a crash-resolution rate (CRR) of $48$\%, compared to $31.5$\% by SWE-agent~\citep{yang2024sweagentagentcomputerinterfacesenable} and $31$\% by Agentless~\citep{xia2024agentlessdemystifyingllmbasedsoftware}, using OpenAI's GPT-4o model.
    Scaling up sampling budget to $10$ trajectories increases \cresearcher{}'s CRR to $54\%$. \cresearcher{} is also robust to model choices, reaching $67\%$ with the newer Gemini 2.5-Flash model.
   Through another experiment on an open-source multimedia software, we show the generalizability of \cresearcher{} and also conduct ablations.
    Our experiments highlight the importance of global context gathering and multi-faceted reasoning for large codebases.
\end{abstract}

\input{sections/introduction}

\input{sections/relatedWork}

\input{sections/methodology}

\input{sections/experimental_setup}

\input{sections/rq}

\input{sections/rq1}

\input{sections/rq2}

\input{sections/rq3}

\input{sections/rq4}

\input{sections/rq5}

\input{sections/rq6}

\input{sections/conclusions}

\bibliography{references}
\bibliographystyle{iclr2026_conference}

\newpage
\appendix

\input{appendix/additional-exp}
\input{appendix/design-details}
\input{appendix/examples}
\input{appendix/ffmpeg}
\input{appendix/example-figs}
\input{appendix/long-context}
\input{appendix/exp-details}
\input{appendix/commit-importance}
\input{appendix/llm-as-judge}
\input{appendix/qualitative_examples}

\end{document}

%% file: math_commands.tex
\usepackage{amsmath,amsfonts,bm}

\def\eqref#1{equation~\ref{#1}}
\def\1{\bm{1}}

\DeclareMathAlphabet{\mathsfit}{\encodingdefault}{\sfdefault}{m}{sl}
\SetMathAlphabet{\mathsfit}{bold}{\encodingdefault}{\sfdefault}{bx}{n}

%% file: sections/introduction.tex
\section{Introduction}%
\label{sec:intro}

Automating coding using Large Language Models (LLMs) and LLM-based agents is a very active area of research. Popular benchmarks like LiveCodeBench~\citep{jain2024livecodebench} and SWE-bench~\citep{jimenez2024swebench} respectively test coding abilities on standalone competitive coding problems and GitHub issues over library or application code. Despite the demonstrated progress of coding agents on these benchmarks, they are yet to scale to complex tasks over an important class of code, \textit{systems code}.

Systems code powers critical and foundational software like operating systems, networking stacks, cloud infrastructure and system utilities. Systems codebases have multiple dimensions of complexity.
Firstly, they are \emph{very large}, with thousands of files and millions of lines of code.
Secondly, systems code often interfaces directly with the hardware and is performance critical. This results in \emph{complex low-level code} (involving pointer manipulations, compile-time macros, etc.) in languages like C/C++, and \emph{global interactions} between different parts of the codebase for concurrency, memory management, maintenance of data-structure invariants, etc.
Finally, foundational systems codebases have \emph{rich development histories} spanning years or even decades, containing contributions by thousands of developers, which are important references on legacy design decisions and code changes.

We consider the problem of generating patches to mitigate crashes reported in systems code.
This is a uniquely challenging setting, as opposed to the SWE-bench~\citep{jimenez2024swebench} like setting tackled by many prior works~\citep{yang2024sweagentagentcomputerinterfacesenable,regym,repograph} using LLMs and SLMs in coding agents.
SWE-bench contains human-written issue descriptions from moderately-sized codebases, which explain the nature of the bug and might indicate which files are likely relevant.
Coding agents~\citep{yang2024sweagentagentcomputerinterfacesenable,openhands} are designed to take advantage of this and quickly navigate the repository to reach the buggy files. 
In fact, an Agentless~\citep{xia2024agentlessdemystifyingllmbasedsoftware} approach, which has a simpler, fixed workflow of localizing the files to edit followed by repair, performs competitively on SWE-bench.
These approaches do not expend much efforts in gathering codebase-wide, global context.
In our setting, the bugs are described by stack traces which are devoid of natural language hints and contain a much larger number of files and functions than an issue description.
Due to the nature of crash reports and the complex global interactions in large systems codebases, multi-step reasoning and context gathering become important.

Automating such complex tasks in systems codebases requires a different type of agents, agents that can \emph{research} about many pieces of context, derived automatically from the large codebase and its massive commit history, \emph{before} making changes.
Recently, \emph{deep research} agents have been developed to solve complex, knowledge-intensive problems that require careful context gathering and multi-step reasoning, before synthesizing the answer. The agents and techniques have mostly focused on long-form document generation or complex question-answering over web contents~\citep{shao2024assisting,openaidr,geminidr,perplexitydr,li2025webthinker,wu2025unfolding} and enterprise data~\citep{anthropicdr,microsoftdr}.
Inspired by these advances, we propose the first deep research agent for code, called \cresearcher{}, and apply it to the problem of generating patches for mitigating crashes reported in systems code.

As shown in Figure~\ref{fig:design}, \cresearcher{} works in three phases:
(1) \analysisphase: Starting with the crash report and the codebase, this phase performs multi-step reasoning over semantics, patterns, and commit history of code. The ``Reasoning Strategies'' block shows the reasoning strategies used. Each reasoning step is followed by invocations of tools (labeled ``Actions'' in Figure~\ref{fig:design}) to gather context over the codebase and its commit history. The information gathered is stored in a context memory and when the agent is able to conclude that it has gathered sufficient context, it moves to the next phase.
(2) \synthesisphase: The \synthesisphase phase uses the crash report, the context memory, and the reasoning trace of the \analysisphase{} phase to filter out irrelevant memory contents. Then, it generates patches, which may edit one or more buggy code snippets from memory, possibly spread across multiple files.
(3) \validationphase: Finally, the \validationphase phase checks if the generated patches prevent the crash from occurring using external tools. A successful patch is presented to the user.

We evaluate the effectiveness of \cresearcher{} on the \kbenchsyz{} benchmark~\citep{kgym}, containing $279$ Linux kernel crashes detected by the Syzkaller fuzzer~\citep{syzkaller}.
This benchmark is challenging because the Linux kernel~\citep{linux} is a canonical example of a systems codebase with complex low-level code and massive size ($75$K files and $28$M lines of code), and has rich development history.
\cresearcher{} resolves $48\%$ of crashes using GPT-4o and $5$ sampled patches, significantly outperforming the two strong and popular baselines of SWE-agent~\citep{yang2024sweagentagentcomputerinterfacesenable} ($31.5\%$) and Agentless~\citep{xia2024agentlessdemystifyingllmbasedsoftware} ($31\%$) in the same setting (and customized for the kernel crash resolution task).
A concurrent work, CrashFixer~\citep{crashfixer}, explores a simpler setting where the agent is provided the ground-truth buggy files to edit (the \emph{assisted} setting), whereas, \cresearcher{} takes only the crash report as input (the \emph{unassisted} setting).

\cresearcher{} benefits from scaling inference compute, reaching $54\%$ with pass@10, and is robust to the choice of model, resolving $67\%$ of crashes with the newer Gemini 2.5-Flash LLM.
It gathers context of high coverage and quality, exploring about $10$ files per trajectory compared to a much smaller number $1.33$ of files explored by SWE-agent. 
We demonstrate the importance of causal analysis over historical commits, a novel feature of \cresearcher{}.
To ensure that the proposed patches do not break existing functionality, we run Linux kernel unit tests on \cresearcher{}'s crash-resolving patches. Though expensive to run, this provides additional validation beyond the crash-reproduction testing in \kbenchsyz{}.
We give evidence of the generalizability of \cresearcher{} by experimenting on an open-source multimedia software, \ffmpeg{}~\citep{ffmpeg}, where it resolves $7/10$ crashes tested.

In summary, we make the following main contributions:\\
\textbf{(1)} We design the first deep research agent for code, \cresearcher{}, capable of handling large systems code and resolving crashes. Recognizing the importance of commit history in systems code, we equip the agent with a tool to efficiently search over commit histories.\\
\textbf{(2)} We evaluate \cresearcher{} on the challenging \kbenchsyz{} benchmark~\citep{kgym} and achieve a crash resolution rate of $54\%$, 
outperforming strong baselines and showing robust performance across model choices, reaching $67\%$ with the newer Gemini 2.5-Flash model.
We also demonstrate its generalizability through experiments on a multimedia software, FFmpeg.\\
\textbf{(3)} Through a comprehensive evaluation, we show (i) how our deep research agent outperforms agents that do not focus on gathering relevant context, (ii) that this advantage persists even if the existing SOTA agent is given higher inference-time compute, and (iii) that reasoning models improve performance significantly if given well-researched context. We thoroughly validate \cresearcher{}'s crash-resolving patches using the Linux kernel unit test-suite in addition to the validation setup in \kbenchsyz{}, providing confidence that they do not break existing functionality. Further ablations show the importance of (i) causal analysis over historical commits and (ii) memory filtering.

%% file: sections/relatedWork.tex
\section{Related work}
\label{sec:related}
The LLM-powered software development subfield has produced several coding agents~\citep{yang2024sweagentagentcomputerinterfacesenable, xia2024agentlessdemystifyingllmbasedsoftware, openhands, autocoderover, wadhwa2024masai}, predominantly evaluated on SWE-bench~\citep{jimenez2024swebench}. SWE-bench focuses on GitHub issues from small to medium-sized Python repositories. However, systems code, the focus of our work, presents unique challenges.
We highlight and contrast key related work in this context. 
A related, but orthogonal line of exploration is long context reasoning. But it has its own challenges, as discussed in Appendix~\ref{app:long-context}.

\textbf{Coding agents}
Agents like SWE-agent~\citep{yang2024sweagentagentcomputerinterfacesenable} or OpenHands~\citep{openhands} use a single ReAct-style~\citep{yao2023react} loop endowed with shell commands or specialized tools for file navigation and editing. 
However, they tend to explore a small number of files per bug, without gathering and reasoning over the relevant codebase-wide context.
AutoCodeRover~\citep{autocoderover} uses tools based on program structure to traverse the codebase (albeit limited to Python code).
It performs explicit localization of the functions/classes to edit using these tools, and those are later repaired.
\cresearcher{} does not explicitly localize the functions to edit; instead it gathers relevant context for patch generation and decides what to edit in the \synthesisphase phase.
Some recent coding agents construct a dependency graph of the repository~\citep{repograph,locagent}, which they then explore using approaches like MCTS~\citep{lingma}. However, such agents are (1) limited to Python code and (2) scale very poorly, making it impractical to use on codebases of the scale of the Linux kernel. \cresearcher{} instead uses simple and scalable tooling to handle such codebases easily.
\cresearcher{} is also the first agent to use causal analysis over historical commits; this is critical to handling subtle bugs introduced by code evolution in long-lived systems codebases.

\textbf{Deep research agents} Deep research is a fast emerging subfield in agentic AI \citep{microsoftdr,openaidr,geminidr,perplexitydr}, to tackle complex, knowledge-intensive tasks, that can take hours or days even for experts. Academic work so far has focussed on long-form document generation \citep{godbole2024analysis,shao2024assisting}, scientific literature review \citep{wu2025unfolding,gottweis2025towards}, and complex question-answering \citep{li2025webthinker,wu2025agentic} based on the web corpus. The key challenges in deep research for such complex tasks include (a) intent disambiguation, (b) exploring multiple solution paths (breadth of exploration), (c) deep exploration (iterative tool interactions and reasoning), and (d) grounding (ensuring that the claims in the response are properly attributed). Most of the aforementioned challenges also apply to our setting. To the best of our knowledge, our work is the first to design and evaluate a deep research strategy for complex bug resolution in large codebases. Most recently, OpenAI's Deep Research model has been integrated with GitHub repos for report generation and QA over codebases \citep{OpenAIDRGH}. However, (a) it does not support agentic tasks like bug fixing, and
(b) their indexing technique does not scale to very large codebases like the Linux kernel, whereas we use scalable tools for search.

\textbf{Automated kernel bug detection and repair} Prior work for detecting Linux kernel bugs includes various types of sanitizers, e.g., Kernel Address Sanitizer (KASAN)~\citep{kasan}, and the Syzkaller kernel fuzzer~\citep{syzkaller}, an unsupervised coverage-guided fuzzer that tries to find inputs to crash the kernel. 
\cresearcher{}, complementary to this, generates patches from crash reports.
We use some traditional software engineering concepts like deviant pattern detection \citep{deviant} and reachability analysis~\citep{nielson2015principles}, but leverage LLMs to scale to large codebases.
As noted earlier, CrashFixer~\citep{crashfixer} targets Linux kernel crashes but assumes that buggy files are known \emph{a priori}. This assumption is unrealistic for large codebases like the Linux kernel. In contrast, \cresearcher{} autonomously locates buggy files using general search tools.

%% file: sections/methodology.tex
\section{Design of \cresearcher{}}%
\label{sec:solution}

Large systems codebases, owing to their critical nature, undergo strict code development and reviewing practices by expert developers. The bugs that still sneak in are subtle and involve violations of global invariants (e.g., a data structure should be accessed only after acquiring a lock) and coding conventions (e.g., use of a specific macro to allocate memory), and unintended side effects caused by past changes. To fix such bugs, an agent needs to gather sufficient context from the codebase and its commit history, before generating hypotheses about the cause of a bug and attempt to fix it. With this insight, we design our deep research agent, \cresearcher{}. As shown in Figure~\ref{fig:design}, \cresearcher{} comprises of three phases: (1) \analysisphase, (2) \synthesisphase and (3) \validationphase.
We present the key details of our design in this section and complement it with implementation details in Appendix~\ref{app:design}.
We also explain an example trajectory of \cresearcher{} in Appendix~\ref{app:examples}.

\subsection{\analysisphase phase}

The \analysisphase phase of \cresearcher{} is responsible for performing deep research to understand the cause of a reported crash.
We equip this phase with (a) actions to efficiently search over the codebase and the commit history and (b) reasoning strategies for code. At each step, the actions taken so far along with their results are stored in a context memory, which is used to construct the prompt.

\subsubsection{Actions to search over codebase and commit history}

We support the following actions: (1) \texttt{search\_definition(sym)}: To search for the definition(s) of the specified symbol, which can be the name of a function, struct, global constant, union or macro and so on. It can be optionally passed a file name to limit the search. (2) \texttt{search\_code(regex)}: To search the codebase for matches to the specified regular expression. This is a simple yet powerful tool, which can be used for searching for any coding pattern such as call to a function, dereferences to a pointer, assignment to a variable and so on. (3) \texttt{search\_commits(regex)}: To search for matches to a regular expression over commit messages and diffs associated with the commits. The regular expression offers expressiveness, e.g., to search for occurrence of a term (``memory leak'') in the commit messages or coding patterns in code changes (diffs). In addition, the agent can invoke (4) \texttt{done} to indicate that it has finished the \analysisphase phase and (5) \texttt{close\_definition(sym)}: To remove the definition of a symbol from the memory if the symbol is deemed irrelevant to the task.

\subsubsection{Reasoning strategies for code}

We ask the agent to explore the codebase to figure out the root cause of a crash and gather sufficient context to propose a fix.
We induce the following reasoning strategies through prompting to guide the exploration of the codebase and its commit history. As shown in Figure~\ref{fig:design}, each reasoning step is followed by one or more actions.
Additionally, we present the agent with a simple scratchpad, where it can add important discoveries for future reference.

\noindent\textbf{Chasing control and data flow chains}
The \emph{control flow}~\citep{nielson2015principles} of a code snippet refers to the functions that are called and the branches in it, including conditional statements, loops, \texttt{goto}s and even conditional compilation macros. Given a crash report and some code, the agent is asked to reason about control flow to understand how execution flows between different functions and how it leads to the crash. Similarly, \emph{data flow}~\citep{nielson2015principles} refers to how the values of variables get passed to different functions and how one variable is used to define another. 
So the agent should also reason about how data flows in the code. 
As a result of this reasoning, the agent may invoke a \texttt{search\_definition(sym)} action to search for the definition of \texttt{sym} if it suspects that \texttt{sym} may have something to do with the buggy behavior and needs more information about \texttt{sym} to confirm or dispel the suspicion. It can also use other actions as suitable, e.g., \texttt{search\_code(x\textbackslash{}s*=)} to look for assignments to a variable named \texttt{x}, with \texttt{\textbackslash{}s*} indicating zero or more whitespaces.

\noindent\textbf{Searching for patterns and anti-patterns}
Traditional software engineering literature thinks of bugs as anomalies -- patterns of code that are deviant~\citep{deviant}. It follows that, to diagnose and understand a bug, one can find frequent patterns in the repository and check if a given piece of code deviates from it. \cresearcher{} reasons about which behavior is common or ``normal'' and which code snippets look anomalous. It can perform a \texttt{search\_code(regex)} action to search for these patterns and anti-patterns using regular expressions. A classic case is checking a pointer for null value after allocation. If the agent notices a missing null check for \texttt{ptr}, it can perform \texttt{search\_code(if\textbackslash{}s*\textbackslash{}(ptr==NULL\textbackslash{}))} to search for null checks throughout the codebase on \texttt{ptr}. Similarly, it can perform \texttt{search\_code(ptr\textbackslash{}s*=.*alloc\textbackslash{}(.*\textbackslash{}))} to search for all allocations to \texttt{ptr} to verify whether other parts of the codebase typically perform a null check or not.

\noindent\textbf{Causal analysis over historical commits}
An interesting and challenging aspect of a codebase that has been in development for a long time, as many foundational systems codebases have, is the rich history of commits. Because of continuous development, it is likely that a new bug has some past commits that can prove helpful in understanding or solving it. Indeed, developers often reference other commits when they come up with patches.
\cresearcher{} reasons about how the codebase has evolved and how that evolution is related to the crash report. It can issue a \texttt{search\_commits(regex)} action to search over past commit messages and diffs. 
For instance, the regular expression \texttt{handle->size|crypto\_fun\textbackslash{}(} matches commits that add or remove a \texttt{handle->size} access, or a call to \texttt{crypto\_fun}.

\noindent\textbf{Iterative process of deep research}
As shown in Figure~\ref{fig:design}, in each reasoning step, \cresearcher{} is asked to decide if it has acquired sufficient context to understand and solve the crash. If yes, it moves to the next phase of synthesizing the patch (Section~\ref{sec:synth}). Initially, the context is empty and it starts its reasoning process by analyzing the contents of the stack trace and the diagnostic information provided as input. 
At each step, the agent evaluates the context accrued so far and decides which lines of exploration to extend by issuing multiple search actions simultaneously.

\subsection{\synthesisphase and \validationphase phases}
\label{sec:synth}

The contents of memory and the reasoning trace of the \analysisphase phase are passed to the \synthesisphase phase, along with the crash report. The \analysisphase phase has the flexibility to follow multiple paths of inquiry simultaneously. It can thus end up collecting information that does not turn out to be relevant, which also happens when a human does research on some topic.
In large codebases, this irrelevant information can be quite large and can overwhelm the prompt.
Thus, the \synthesisphase phase first filters the memory and discards (action, result) pairs that are deemed irrelevant to the task of fixing the crash. The agent then uses the filtered information to generate a hypothesis about the nature of the bug and a potential remedy, and the corresponding patch. Finally, in the \validationphase phase, the patch is applied to the codebase, and the codebase is compiled. The reproducer program that had originally caused a crash is run. If the crash is reproduced, the patch is rejected. If not, it is accepted.

%% file: sections/experimental_setup.tex
\section{Experimental setup}%
\label{sec:experiments}

\noindent\textbf{Benchmarks} We use a thoroughly validated, reproducible subset of \textbf{200 instances} from the \kbenchsyz{} benchmark~\citep{kgym} of $279$ Linux kernel crashes found by the Syzkaller fuzzer~\citep{syzkaller}.
Each instance in the benchmark consists of (1) a reproducer file, containing the user-space program that triggers the crash, (2) the ground-truth commit that fixed the bug, 
and (3) the crash report at the parent commit of the fix commit (we run all tools at this parent commit).
To show generalizability, we also evaluated \cresearcher{} on $10$ recent crashes of an open-source multimedia software, FFmpeg~\citep{ffmpeg}.
More details about the \kbenchsyz{} benchmark and the \ffmpeg{} dataset are in Appendix~\ref{app:details} and Appendix~\ref{app:ffmpeg} respectively.

\noindent\textbf{Evaluation metrics}
We compute Pass@k~(\textbf{P@$k$}) defined as $\textbf{P@$k$} = 1$, if applying at least one of the $k$ candidate patches generated by the tool prevents the crash, 
or $\textbf{P@$k$} = 0$ otherwise. We report \textbf{(1)} \textbf{C}rash \textbf{R}esolution \textbf{R}ate~(\textbf{CRR}) which is average \textbf{P}@$k$, \textbf{(2)} average \textbf{recall}, i.e., the fraction of files modified in the ground-truth commit (\emph{the ground-truth buggy files}) in the set of files edited by the agent, averaged over the $k$ candidate patches, and 
\textbf{(3)} the percentage of candidate patches where \textbf{All}, \textbf{Any} or \textbf{None} of the ground-truth buggy files are edited. 
When a tool does not produce a patch (e.g., it runs out of LLM call budget), the set of edited files is assumed to be empty. All the metrics are averaged over the $200$ instances in the benchmark. 

\noindent\textbf{Baselines}
We evaluate \textbf{\cresearcher{} in the \emph{unassisted} setting} (i.e., the ground-truth buggy files that are part of the fix commits are not divulged to the tool) and compare it against the following baselines:
\textbf{(1)} \textbf{o1}~\citep{o1} \textbf{and GPT-4o}~\citep{gpt4o} \textbf{in the \emph{assisted} setting}, i.e., we give the ground-truth files that are part of the fix commits and the crash report as input. We prompt the model to generate a hypothesis about the crash's cause and a patch.
\textbf{(2)} \textbf{o1 and GPT-4o in the \emph{stack context} setting}, where we give the contents of the files mentioned in the crash report as input besides the crash report. 
\textbf{(3)} \textbf{SWE-agent 1.0}~\citep{yang2024sweagentagentcomputerinterfacesenable}, a SOTA coding agent on the SWE-bench benchmark, \textbf{in the \emph{unassisted} setting}.
For fairness, we add a Linux kernel-specific example trajectory and background about the kernel to its prompts.
We sample $k$ (for Pass@$k$) SWE-agent trajectories independently using a temperature of $0.6$.
\textbf{(4)} \textbf{Agentless}~\citep{xia2024agentlessdemystifyingllmbasedsoftware}, another top coding agent on the SWE-bench benchmark that uses a fixed workflow of localization followed by repair, \textbf{in the \emph{unassisted} setting}. We add background about the kernel to its prompts and sample $k$ (for Pass@$k$) patches using the same LLM (GPT-4o) as \cresearcher{} and SWE-agent. 
\textbf{(5)} \textbf{CrashFixer}~\citep{crashfixer}, state-of-the-art agent for Linux kernel crash resolution, \textbf{in the \textit{assisted} setting}, as it directly uses the ground-truth files to generate patches. If a patch fails to build or crashes with the ground-truth reproducer, it iteratively refines it using the respective error messages. We report their results with the Gemini 1.5-Pro-002 model (v. 2024-09-24) and more recent results with the Gemini 2.5-Pro model (v. 2025-06-17).

\noindent\textbf{Implementation, hyperparameters}
We employ \textbf{GPT-4o} (v. 2024-08-06) for \cresearcher{} and for the competing tools. We also experiment with \textbf{o1} (v. 2024-12-17) in the \synthesisphase{} phase of \cresearcher{}, and,
to show that \cresearcher{}'s design is not tied to the OpenAI family of models, the newer \textbf{Gemini 2.5-Flash} (v. 2025-06-17) for both the \analysisphase{} and \synthesisphase{} phases.
All our experiments have a context length limit of $50$K tokens.
In the \analysisphase{} phase, we use a \textbf{temperature} of $0.6$ and independently sample $k$ trajectories.
For the \synthesisphase{} phase, we sample with increasing temperatures ($0$, $0.3$, $0.6$) until the agent produces a correctly-formatted patch, with a maximum of $3$ attempts.
We allow all tools a budget of at most \textbf{\maxcalls{}} LLM calls to generate a single patch.
Please refer to Appendix~\ref{app:details} for details about the \textbf{crash reproduction setup}, the \textbf{prompts}, the \textbf{compute resources} and the \textbf{configurations} for \cresearcher{} and the baselines.

%% file: sections/rq.tex
\section{Experimental results}%
\label{sec:evaluation}
We evaluate \cresearcher{} across six dimensions.
First, we compare its crash resolution ability against state-of-the-art coding agents and baselines.
We then analyze the context gathering capabilities of different tools by studying (a) whether they are able to find the files that need to be edited, and (b) the coverage and quality of additional global context gathered.
Then, we assess the impact of two design choices, historical commit analysis and context filtering, on \cresearcher{}'s performance.
Finally, we further validate \cresearcher{}’s crash-resolving patches with unit tests and a qualitative analysis, and show its generalizability to another codebase.

%% file: sections/rq1.tex
\begin{table}[t]
\centering
\caption{Crash resolution rate (CRR) for different tools on the \kbenchsyz{} benchmark ($200$ bugs). LLMs used by the tools are in parentheses. $^*$Results are from \citet{crashfixer}, out of $279$ bugs.
}
\label{tab:rq1-main}
\vspace*{-2mm}
\scalebox{0.95}{
\begin{tabular}{ccccc}
\cmidrule{1-5}
\textbf{Setting} & \textbf{Tool} & \textbf{Max calls}& \textbf{P@k} & \textbf{CRR (\%)} \\
\cmidrule{1-5}
\multirow{3}{*}{\rotatebox[origin=c]{0}{\centering Assisted}} 
  & GPT-4o           & $1$ & P@$5$ & $36.00$ \\
  & o1                       & $1$ & P@$5$  & $51.00$ \\
  & CrashFixer~(Gemini 1.5 Pro-002)$^*$ & $\geq 4$ & P@$16$ & $49.22^*$ \\
  & CrashFixer~(Gemini 2.5 Pro) & $\geq 4$ & P@$16$ & $\mathbf{70.00}$ \\
\cmidrule{1-5}
\multirow{2}{*}{\rotatebox[origin=c]{0}{\centering Stack context}} 
  & GPT-4o & $1$ & P@$5$  & $29.50$ \\
  & o1 & $1$ & P@$5$  & $\mathbf{40.00}$ \\
\cmidrule{1-5}
\multirow{3}{*}{\rotatebox[origin=c]{0}{\centering Unassisted}} 
  & Agentless (GPT-4o) & $4$ & P@$5$  & $31.00$ \\
  & SWE-agent (GPT-4o) & $15$ & P@$5$  & $31.50$ \\
  & \textbf{\cresearcher{} (GPT-4o)} & $15$ & P@$5$  & $48.00$ \\
  & \textbf{\cresearcher{} (GPT-4o + o1)} & $15$ & P@$5$  & $58.00$ \\
  & \textbf{\cresearcher{} (Gemini 2.5-Flash)} & $15$ & P@$5$  & $\mathbf{67.00}$ \\
\cmidrule{1-5}
\multirow{4}{*}{\rotatebox[origin=c]{0}{\centering \makecell{Unassisted \\ + Scaled} }} 
  & SWE-agent (GPT-4o) & $30$ & P@$5$  & $32.00$ \\
  & \textbf{\cresearcher{} (GPT-4o)} & $30$ & P@$5$  & $47.50$ \\
  & SWE-agent (GPT-4o) & $15$ & P@$10$ & $37.50$ \\
  & \textbf{\cresearcher{} (GPT-4o)} & $15$ & P@$10$  & $\mathbf{54.00}$ \\
\cmidrule{1-5}
\end{tabular}
}
\end{table}

\subsection{RQ1: How effective are different tools at resolving Linux kernel crashes?}%
\label{sec:rq1}

Our main results are presented in Table~\ref{tab:rq1-main}, organized by setting, namely, assisted, stack context, unassisted, and unassisted + test-time scaled (Section~\ref{sec:experiments}).

\noindent\textbf{1) The assisted setting baselines show strong performance, but under unrealistic assumptions.}
Given the ground-truth buggy files, LLMs like GPT-4o (CRR of $36\%$) are quite capable of resolving crashes.
A reasoning model like o1 is significantly better (CRR of $51\%$).
Adding iterative feedback from the compiler and the crash reproduction setup, CrashFixer achieves $49.22\%$ CRR with an older Gemini model and $70\%$ CRR with a newer Gemini reasoning model
using P@$16$ with at least $4$ \maxcalls{}.

\noindent\textbf{2) The stack context setting reveals the difficulty of the practical unassisted setting.}
The assumption that an oracle can tell us exactly which files need to be edited is impractical.
The gap between the assisted (idealistic) setting and the practical unassisted setting is highlighted by the simple but effective stack context setting
where models are given the contents of all the files mentioned in the crash report (truncated to the context length limit).
This is a strong baseline because all the ground-truth buggy files are present in the crash report for {$74.50\%$} crashes in our dataset.
o1 achieves a CRR of $40\%$, which is impressive, but $11\%$ lower than its performance in the assisted setting.

\noindent\textbf{3) \cresearcher{} consistently outperforms baselines in the unassisted setting.}
Fixing the LLM as GPT-4o, \cresearcher{} achieves a CRR of $48\%$, significantly outperforming the SWE-agent and Agentless baselines, both the stack context baselines, and even the assisted GPT-4o baseline.
This indicates that the context gathered by \cresearcher{} is much more effective than giving file contents based on the crash report, or using multi-step hierarchical localization (as done by Agentless), and is even better than directly giving all the contents of the files to be edited.
The context gathered by GPT-4o during \analysisphase{} can be better utilized by the reasoning model o1 during \synthesisphase{}, increasing \cresearcher{}'s CRR to $58\%$.
This further increases to $67\%$ using the newer (albeit not as advanced as Gemini 2.5-Pro) model, Gemini 2.5-Flash, for both phases, indicating that \cresearcher{}'s design is not tied to a specific LLM family.

\noindent\textbf{4) Increasing the number of trajectories helps, while making them longer does not.}
We examine how scaling the total inference budget (\maxcalls{} $\times$ num trajectories $k$) impacts the performance of \cresearcher{} and SWE-agent.
Doubling the \maxcalls{} budget, i.e., making the trajectories of the agents longer, has a negligible effect on the CRR, whereas
increasing the number of trajectories sampled improves SWE-agent's CRR to $37.50\%$ and \cresearcher{} (GPT-4o)'s CRR to $54.00\%$.

%% file: sections/rq2.tex
\subsection{RQ2: Do the tool-edited files match those modified in developer fixes?}%
\label{sec:rq2}

The information needed by a developer to fix a crash can be divided into (1) the code to be edited and (2) additional context. We study (1) here by examining whether tools edit the same files as developers, and study (2) in the next section. Since buggy files are already provided in the assisted setting, we focus on the stack context and unassisted settings. Full results appear in Table~\ref{tab:rq2-main}. 

\begin{table}[t]
\centering
\caption{Average recall and All/Any/None percentages (metrics defined in Section~\ref{sec:experiments}) for different tools. LLMs used by the tools are in parentheses.}
\label{tab:rq2-main}
\begin{tabular}{@{}c@{}ccccc@{}}
\cmidrule{1-6}
\textbf{Setting} & \textbf{Tool} & \makecell{\textbf{Max} \\ \textbf{calls}} & \textbf{P@k} & \makecell{\textbf{Avg.}\\\textbf{Recall}} & \makecell{\textbf{All/Any/None} \\ \textbf{(\%)}} \\
\cmidrule{1-6}
\multirow{2}{*}{\rotatebox[origin=c]{0}{\centering Stack context}}
& GPT-4o & $1$ & P@$5$  & $\mathbf{0.46}$ & $\mathbf{42.5}/7.8/49.7$ \\
& o1 & $1$ & P@$5$  & ${0.43}$ & $39.7/7.7/52.6$ \\
\cmidrule{1-6}
\multirow{3}{*}{\rotatebox[origin=c]{0}{\centering Unassisted}} 
  & Agentless (GPT-4o) & $4$ & P@$5$ & $0.49$ & $46.4 / 7.7 / 45.9$ \\
  & SWE-agent (GPT-4o) & $15$ & P@$5$ & $0.37$ & $35.1/5.6/59.3$ \\
  & \textbf{\cresearcher{} (GPT-4o)} & $15$ & P@$5$ & $0.51$ & $48.2 / 7.8 / 44.0$ \\
   & \textbf{\cresearcher{} (GPT-4o + o1)} & $15$ & P@$5$  & $0.53$ & $49.9/7.6/42.4$ \\
       &\textbf{\cresearcher{} (Gemini 2.5-Flash)} & $15$ & P@$5$  & $\mathbf{0.56}$ & $\mathbf{52.1}/9.7/38.2$ \\
\cmidrule{1-6}
\multirow{4}{*}{\rotatebox[origin=c]{0}{\centering \makecell{Unassisted \\ + Scaled} }} 
  & SWE-agent (GPT-4o) & $30$ & P@$5$  & $0.40$ & $37.9/6.4/55.7$ \\
  & \textbf{\cresearcher{} (GPT-4o)} & $30$ & P@$5$  & $\mathbf{0.53}$ & $\mathbf{49.5}/8.0/42.5$ \\
  & SWE-agent (GPT-4o) & $15$ & P@$10$ & $0.36$ & $34.3/5.5/60.2$ \\
  & \textbf{\cresearcher{} (GPT-4o)} & $15$ & P@$10$  & $0.51$ & $47.8/7.5/44.7$ \\
\cmidrule{1-6}
\end{tabular}
\end{table}

\textbf{\cresearcher{} achieves the highest recall across all baselines}, editing all ground-truth buggy files in nearly half of candidate patches and at least one in another $\sim8\%$. ``Recall'' here measures the fraction of ground-truth files in those \emph{edited}, not merely \emph{explored}, making it notable that \cresearcher{} edits on average only $1.1$ files while exploring $10$ (Section~\ref{sec:rq3}). Agentless attains recall comparable to \cresearcher{} and much higher than SWE-agent, but resolves far fewer crashes. \textbf{This contrast underscores the importance of the global context gathered by \cresearcher{}, beyond just the code being edited.}
Finally, scaling test-time compute (via increasing \maxcalls{} or P@$k$) for \cresearcher{} and SWE-agent preserves the overlap between edited and ground-truth files, and using the newer Gemini 2.5-Flash for \cresearcher{} further improves its overlap.

%% file: sections/rq3.tex
\subsection{RQ3: How effective is context gathering for resolving kernel crashes?}%
\label{sec:rq3}

\begin{figure}[t]
    \centering
    \includegraphics[width=0.45\linewidth,scale=0.4,keepaspectratio]{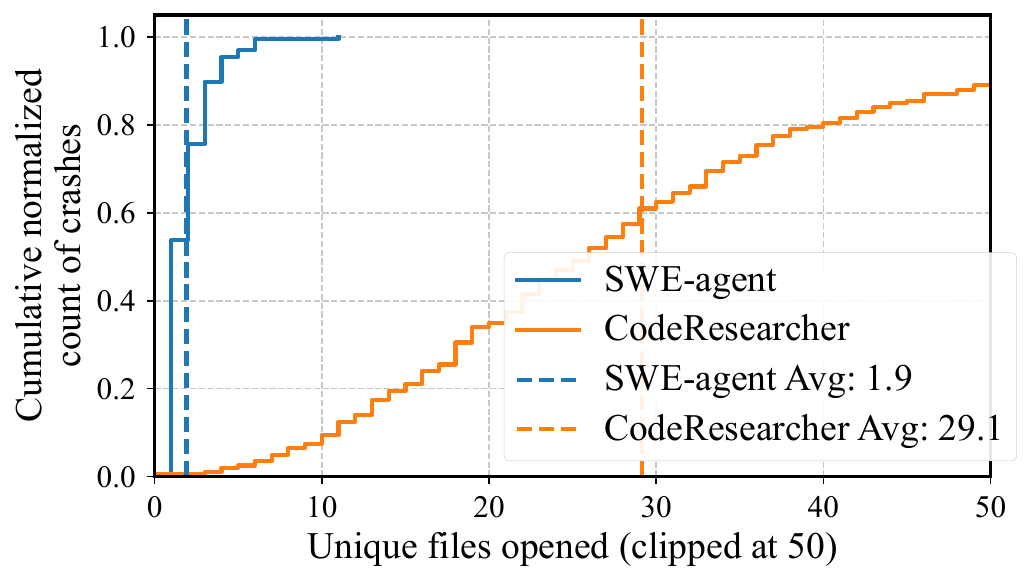}
    \caption{Files explored for each crash (summed over 5 trajectories).}
    \label{fig:histogram}
\end{figure}
From Tables~\ref{tab:rq1-main} and \ref{tab:rq2-main}, \cresearcher{} (GPT-4o) significantly outperforms baselines that do not gather additional context but have high recall.
This shows that gathering global context (instead of just localizing the code to edit) can drastically improve kernel crash resolution.
Agentless, which does not gather any context, performs poorly, while
SWE-agent does gather context from the codebase, yet it performs much worse than \cresearcher{}.
Below, we investigate this further:

\noindent\textbf{1) Coverage of the context gathered:}
Figure~\ref{fig:histogram} shows the distribution of the number of unique files read across the $5$ trajectories by \cresearcher{} and SWE-Agent (GPT-4o, P@$5$, $15$ \maxcalls{}).
\cresearcher{} performs deep research over the codebase, reading $29.13$ unique files across $5$ top-level directories on average for each crash.
In stark contrast, SWE-agent reads only $1.91$ files on average for each crash.
When averaged by trajectory, \cresearcher{} explores $10$ unique files compared to only $1.33$ files explored by SWE-agent.

\noindent\textbf{2) Overlap with developer-referenced context:}
We use LLM-as-judge to determine the overlap of the context gathered by \cresearcher{}~(GPT-4o) and SWE-agent with the context mentioned by the developer in the fix commit message (details in Appendix~\ref{app:llm-as-judge}).
This context overlap is $54.18\%$ (over candidate patches) for SWE-agent compared to $63.7\%$ for \cresearcher{}, suggesting that \cresearcher{} does a much better job of identifying relevant context that the developer explicitly relied on when making the fix.

\noindent\textbf{3) Context quality when both edit all the ground-truth modified files:}
To isolate the impact of the gathered context,
we consider the subset of $90$ crashes, where both \cresearcher{} and SWE-agent (using the same GPT-4o model) edit \textit{all} the ground-truth files in \textit{at least one candidate patch} generated by each tool.
We can thus attribute their success (or failure) on this subset to the context gathered.
\cresearcher{} resolves $55/90 = 61.10\%$ of crashes in this subset, while SWE-agent resolves only $34/90 = 37.78\%$ (discounting crash-resolving patches from each tool that do not edit \textit{all} the ground-truth files).
\emph{Taken together, the three observations show that not only does \cresearcher{} gather more context than SWE-agent, it also gathers higher quality context.}

%% file: sections/rq4.tex
\subsection{RQ4: How important are historical commit analysis and context filtering?}
\label{sec:rq4}

\noindent\textbf{Commit history analysis}
\cresearcher{} is the first agent to explicitly leverage the rich development history of codebases.
In this ablation, we run it without the \texttt{search\_commits} action on the set of $96$ bugs that were successfully resolved by \cresearcher{}~(GPT-4o, Pass@$5$, $15$ \maxcalls{}). 
Table~\ref{tab:rq4-main} (Appendix~\ref{app:additional-exp}) shows that removing the \texttt{search\_commits} action leads to a $10\%$ drop in the crash resolution rate,
and decreases the ability to edit the ground-truth modified files.
This highlights that the \texttt{search\_commits} action plays a crucial role 
in context gathering and localization.
Notably, for the example in Appendix~\ref{app:commit-importance}, we also observe that \cresearcher{} navigates to the \emph{same} buggy commit that originally introduced the bug being repaired.

\noindent\textbf{Context filtering}
As mentioned in Section~\ref{sec:synth}, the \analysisphase{} phase often gathers large amounts of irrelevant context, especially in big codebases like the Linux kernel.
The \synthesisphase{} phase filters this memory before synthesizing a patch.
We provide two pieces of evidence for the importance of this filtering.
First, the average memory length (capped at $50K$) across all \cresearcher{} (GPT-4o, P@$5$, $15$ \maxcalls{}) trajectories  dropped from $21,557$ tokens to $7,797$ tokens after filtering.
Since irrelevant tokens hurt LLM reasoning, this reduction should help performance.
Second, in an ablation on $20$ randomly sampled crashes ($10$ resolved, $10$ unresolved), disabling filtering reduced resolved crashes from $10$ to $8$, average recall from $0.41$ to $0.35$, and All/Any/None from $34.0/16.0/50.0$ to $29.0/15.0/56.0$.
This shows the importance of filtering in \cresearcher{}'s performance.

%% file: sections/rq5.tex
\subsection{RQ5: How robust are \cresearcher{}'s patches?}
\label{sec:rq5}

\noindent\textbf{Kernel unit tests}
In addition to extensive testing (for $10$ minutes on $4$ machines with $8$ parallel processes) for crash-resolution per the \kbenchsyz{} setup, we run kernel unit tests to check if the proposed patches break existing functionality.
For each of the $116$ crashes resolved by \cresearcher{} (GPT-4o+o1, P@$5$, $15$ \maxcalls{}), we selected one crash-resolving patch and ran KUnit~\citep{kunit} tests on it.
For a total of $28$ crashes, either KUnit was not present in the kernel source code version on which the crash was reported or it did not support the required setup.
For the remaining $88$ crashes, all the KUnit tests had a status of either \texttt{PASS} or \texttt{SKIP} (some hardware-specific tests are skipped depending on the machine requirements). On average, only $\sim 13$ tests were skipped for a patch while $\sim 210$ tests were passed with \emph{no unit test failures reported}.

\noindent\textbf{Qualitative analysis}
While perusing the crash-resolving patches, we came across the following types of patches. Examples for each category (with explanations) are in Listings~\ref{lst:jfs_semantic}-\ref{lst:qrtr_misfix}, Appendix~\ref{app:qualexamples}.
\textbf{(1) Accurate} patches correctly identify and fix the root cause of the crash, closely resembling the developer solution.
\textbf{(2) Overspecialized} patches successfully prevent the crash but may be overspecialized.
\textbf{(3) Incomplete} patches correctly identify the problem area and approach, but may not be complete. 
They provide debugging insights and could accelerate the path to a proper fix.
\textbf{(4) Inaccurate} patches offer a plausible way to resolve the crash, but differ from the developer fix.

%% file: sections/rq6.tex
\subsection{RQ6: Does \cresearcher{} generalize to other systems codebases?}%
\label{sec:rq6}
To demonstrate that \cresearcher{} generalizes with a little effort to other codebases, we experiment with crash resolution in the \ffmpeg{}~\citep{ffmpeg} codebase, a leading open-source multimedia framework. We build a small dataset of $10$ recent security-related crash vulnerabilities (which are assigned the top priority) reported by OSS-Fuzz~\citep{ossfuzz}, an automated fuzzing service for open-source projects. Full details of the codebase, dataset construction, and reproduction steps are given in Appendix~\ref{app:ffmpeg}.
We run \cresearcher{} with the same core prompts as for the Linux kernel. 
In the unassisted setting (\analysisphase{} with GPT-4o, \synthesisphase{} with o1, \maxcalls{} = $15$), \cresearcher{} \textbf{resolves $7$ of $10$ crashes} at Pass@$1$. 
It achieves an \textbf{average recall of $0.78$, editing all the ground-truth files in $7$ crashes and none in $2$} (excluding one case without a known fix). 
While \ffmpeg{} crashes are typically not as complex as Linux kernel crashes, our results show that \cresearcher{}’s techniques generalize easily and effectively to other systems codebases.

%% file: sections/conclusions.tex
\section{Conclusions, limitations, and future work}
\label{sec:conclusion}
In this work, we extend coding agents to deep research scenarios arising in resolving complex issues in large systems codebases.
We (a) leverage the rich development history in the codebases (commits), and (b) design effective deep exploration strategies for gathering the rich context often needed to root-cause and patch code crashes.
We establish state-of-the-art results on the latest and challenging benchmark of Linux kernel crashes, thoroughly validate our results, perform ablations, and show the generalizability of our approach.
Our work currently targets the crash resolution problem, but there are other equally important problems faced by systems software such as slow response times, excessive resource usage and flakiness. It remains to be seen if our deep research strategy could be applied to these scenarios. Deep research for code is a new subfield of agentic AI and we intend to explore novel usecases and strategies beyond the ones presented in the paper.

%% file: appendix/additional-exp.tex
\section{Additional experimental results}
\label{app:additional-exp}

\paragraph{Table of results for Section~\ref{sec:rq4}}
Table~\ref{tab:rq4-main} shows the results for the \texttt{search\_commits} ablation on the set of $96$ bugs that were successfully resolved by \cresearcher{} (GPT-4o, Pass@$5$, $15$ \maxcalls{}).

\begin{table}[htp]
\centering
\caption{Importance of causal analysis of past commits on 96 bugs resolved by \cresearcher{}.}
\label{tab:rq4-main}
\begin{tabular}{ccccccc}
\cmidrule{1-7}
& \textbf{Tool} & \makecell{\textbf{Max} \\ \textbf{Calls}} & \textbf{P@k} & \textbf{CRR(\%)} & \makecell{\textbf{Avg.}\\\textbf{Recall}} & \textbf{All/Any/None(\%)}\\
\cmidrule{1-7}
& \textbf{\cresearcher{} (GPT-4o)} & $15$     & P@5  & $\mathbf{48.00}$ & $\mathbf{0.51}$ & $\mathbf{48.2} / 7.8 / 44.0$ \\
& \textsc{w/o} \texttt{search\_commits} & $15$     & P@5  & $38.00$ & $0.33$ & $32.6/2.4/65.0$ \\
\cmidrule{1-7}
\end{tabular}
\begin{flushleft}
\footnotesize{$^1$ We do this ablation only on the 96 bugs resolved by \cresearcher{}~(GPT-4o, Pass@$5$, $15$ \maxcalls{}).}
\end{flushleft}
\end{table}

\begin{figure}[htp]
    \centering
    \includegraphics[scale=0.4,keepaspectratio]{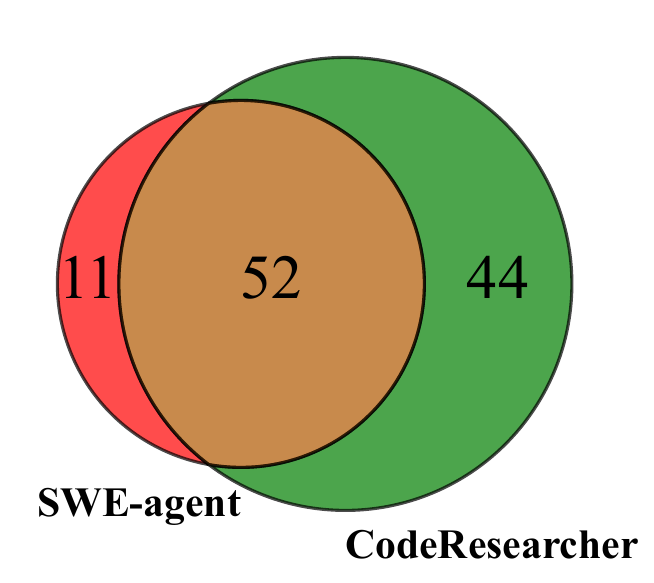}
    \caption{Overlap of crashes resolved by SWE-agent and \cresearcher{} (both at GPT-4o, P@$5$, $15$ max calls)}
    \label{fig:venn}
\end{figure}
\paragraph{Subset analysis of the crashes resolved by \cresearcher{} and SWE-agent}
Figure~\ref{fig:venn} shows a venn diagram of the crashes resolved by \cresearcher{} and SWE-agent in the same configuration (GPT-4o,P@$5$,$15$ \maxcalls{}).
It shows that \cresearcher{} is able to solve most of the crashes that SWE-agent resolves, while also resolving a significant number of additional crashes.

%% file: appendix/design-details.tex
\section{\cresearcher{} implementation details} \label{app:design}

The full implementation can be found in the \texttt{cresearcher} folder in the supplementary material. We explain a few key points here.

\paragraph{Context memory}
\cresearcher{} stores all the context collected so far in a context memory. The memory contains a mapping of file path to a list of symbol definitions (further containing fields like symbol name, start position, end position, body, etc.). It also has search queries and results, storing a list of (query, results) pairs where results is itself a list of results. Please see the definition of the \texttt{class GlobalCtxAgentState} in the file \texttt{cresearcher/utils/types.py} in our supplementary material. In Figure~\ref{fig:complete_example_crash_report_step1} (Appendix~\ref{app:examples}), the context shown provides an example of the memory.
We use the memory to form the prompt for each step.

\paragraph{Prompts}
The system and user prompts for the \analysisphase{} and \synthesisphase{} phases (for both filtering and patch generation) are provided in the \texttt{prompts} folder of our supplementary material. They use a \texttt{prompt\_preamble} and \texttt{prompt\_analysis\_examples} that vary based on the codebase (e.g., Linux kernel or FFmpeg). Those can be found in the files \texttt{config/kBenchSyz.yaml} and \texttt{config/ffmpeg.yaml} respectively in our supplementary material.
We use the context memory to form the prompt for each step. We show all the open symbol definitions (results of \texttt{search\_definition} actions), however we only show the results of queries (\texttt{search\_code} and \texttt{search\_commits}) from the previous step instead of showing them for all past steps. This is because the results of these queries typically have lower signal to noise ratio and commits can be quite large so they threaten to overwhelm the context. The conversation trajectory is also passed to the LLM, but in case it exceeds our limit of $50K$, we remove intermediate messages as necessary (i.e., the first message containing the crash report and the last message are always kept, and intermediate messages are included as much as possible).
In the filtering step (please see \texttt{generateGlobalCtxAgentSelectionPrompt} in \texttt{cresearcher/globalContext.py} in the supplementary), we prioritize fitting symbol definitions first before we try fitting other search queries and results into our context length limit.

\paragraph{Implementation of the search actions}
We implement the \texttt{search\_definition(sym)} action using the \texttt{ctags}~\citep{ctags} tool to generate (and read) an index file of language objects found in source files for programming languages.
The index file is constructed once at the start of \cresearcher{}'s run, usually taking a few minutes for the Linux kernel codebase, and is used throughout the \analysisphase{} trajectory.
Whenever we show a symbol definition in the prompt, for each line of code that is mentioned in the crash report, we additionally add an annotation (as a C-style comment at the end of the line) saying that this line is important.
For \texttt{search\_code(regex)}, we use the \texttt{git grep -E} command to search over all the tracked files in the codebase and show $2$ lines of context before and after each matching line.
Finally, for \texttt{search\_commits(regex)}, we use the \texttt{git log -E -G} and \texttt{git log -E --grep} commands to search over historical commits matching in the code changes and commit messages, respectively.
The message and patch of each relevant commit are returned as output, truncated to a maximum of $100$ lines.
Each action can return a maximum of $5$ results.
To make these searches over an extremely large repository faster, we progressively search over the files of the symbol definitions present in context memory, then those mentioned in the crash report, then those in the kernel subsystems of the bug, and finally all the files in the codebase.
This prioritization strategy allows us to use a timeout of $60$ seconds for the \texttt{git log} commands (which usually take the longest time) while still getting relevant results in a large number of cases.

\paragraph{Scalability of search tooling}
We now give some experimental evidence to substantiate our point in Section~\ref{sec:related} that existing coding agents that construct repository dependency graphs, like LocAgent~\citep{locagent}, RepoGraph~\citep{repograph} and Lingma Agent~\citep{lingma}, scale poorly to large codebases.

We ran the LocAgent and RepoGraph graph constructions for the \texttt{sympy} repository from SWE-bench that contains $\sim 433 K$ lines of Python code. LocAgent took $764$ seconds. RepoGraph errored out after $44.3$ seconds and the progress bar showed \texttt{20/1584 [00:24<31:42, 1.22s/it]}, indicating that it would have taken another $30$ minutes to complete. In contrast, constructing the \texttt{ctags} index used by \cresearcher{} doesn't require any dependency analysis and took 0.76 seconds for the same repository.

Given the times above and the fact that the Linux kernel has $\sim 28 M$ lines of code, even if one were to change their tools to support multiple languages, it would be infeasible to construct dependency graphs in reasonable time. On the other hand, constructing the \texttt{ctags} index for the Linux kernel takes only $74$ seconds.

%% file: appendix/examples.tex
\section{Example of agent trajectory}%
\label{app:examples}

In this section, we show an example \analysisphase{} trajectory generated by \cresearcher{} while solving a kernel crash from the dataset\footnote{Bug in syzkaller: \url{https://syzkaller.appspot.com/bug?id=35dd38010e383ee37d27b39f6b50eb25b4c92dd2}}.
We explain the example here and provide the complete trajectory for reference in Figures~\ref{fig:complete_example_crash_report}-\ref{fig:complete_example_crash_report_step5} of Appendix~\ref{app:example-figs}.
The truncated crash report that initiates the investigation of the example is shown in Figure~\ref{fig:complete_example_crash_report}.

The trajectory begins in Figure \ref{fig:complete_example_crash_report_step0}. \cresearcher{} identifies a warning in \verb|smsusb_term_device| related to \verb|__flush_work| and forms an initial hypothesis about workqueue synchronization issues.
The agent then  explores the codebase by examining the \verb|__flush_work| implementation and analyzing the \verb|smsusb_stop_streaming| function's synchronous URB cancellation pattern in Figure \ref{fig:complete_example_crash_report_step1}, while also searching for relevant commit history to identify potential race conditions.

In Figures \ref{fig:complete_example_crash_report_step2}–\ref{fig:complete_example_crash_report_step4}, \cresearcher{} traces prior fixes for related bugs in \verb|smsusb_term_device|,  inspects synchronization behavior in \verb|cancel_work_sync| and \verb|__cancel_work_timer|. It uncovers a critical misstep where work item initialization via \verb|INIT_WORK| in \verb|smsusb_onresponse| can be bypassed, resulting in NULL \verb|work->func| pointers that explain the observed warning.

Finally, in Figure \ref{fig:complete_example_crash_report_step5}, \cresearcher{} confirms the root cause: though \verb|smsusb_onresponse| correctly initializes work structures with \verb|INIT_WORK|, it is not sufficient. If \verb|smsusb_stop_streaming| is called before any URB completion occurs, the system attempts to cancel uninitialized work items, triggering warnings in \verb|__flush_work| when it encounters NULL function pointers.

%% file: appendix/ffmpeg.tex
\section{\ffmpeg{}: experimental details}%
\label{app:ffmpeg}

\paragraph{Background} 
\ffmpeg{} is a leading open-source multimedia framework that supports decoding, encoding, transcoding, muxing, demuxing, streaming, filtering, and playback of virtually all existing media formats.
Since it needs to handle a wide range of formats, from very old to the cutting edge, low-level data manipulation is common in the codebase. 
As of May 2025, \ffmpeg{} has $\sim 4.8$K files and $\sim 1.46$M lines of code, primarily in C / C++, with some handwritten assembly for performance.

\paragraph{Dataset}
We use vulnerabilities discovered by the OSS-Fuzz service~\citep{ossfuzz} that runs fuzzers on various open source projects and creates alerts for the bugs detected.
We focus on security issues, which are assigned the top priority by OSS-Fuzz.
These include heap-based buffer overflows, stack-based buffer overflows, use-after-frees, etc.
We build a small dataset of $10$ \ffmpeg{} crashes, taking the $11$ most recent crashes (as of May $14$, $2025$) that have been verified as fixed and skipping $1$ that we could not validate.
\footnote{\url{https://issues.oss-fuzz.com/issues?q=project:ffmpeg\%20type:vulnerability\%20status:verified&s=modified_time:desc&p=1}}
We use the instructions recommended by OSS-Fuzz for building \ffmpeg{} and testing whether a crash reproduces.
\footnote{\url{https://google.github.io/oss-fuzz/advanced-topics/reproducing/}}
The dataset contains the commit at which OSS-Fuzz found the crash, a reproducer file that triggered the crash, and the crash report that we generated by reproducing the crash (the crash reports found by OSS-Fuzz are not publicly visible).
We make the dataset available in the \texttt{data/ffmpeg} folder in our supplementary material.

\paragraph{Reproduction}
To run \cresearcher{} on these crashes, we keep the same core prompts (which can be found in the \texttt{prompts} folder in the supplementary material), adding a one-paragraph preamble about \ffmpeg{} and replacing the few-shot examples for the Linux kernel with corresponding ones for \ffmpeg{}.
The preamble and few-shot examples are in the file \texttt{config/ffmpeg.yaml} in our supplementary material.
The \texttt{README} also contains instructions for running \cresearcher{} on an \ffmpeg{} crash.

%% file: appendix/example-figs.tex
\section{Example agent trajectory: figures}
\label{app:example-figs}

\begin{figure}[H]
    \centering
    \includegraphics[width=\linewidth]{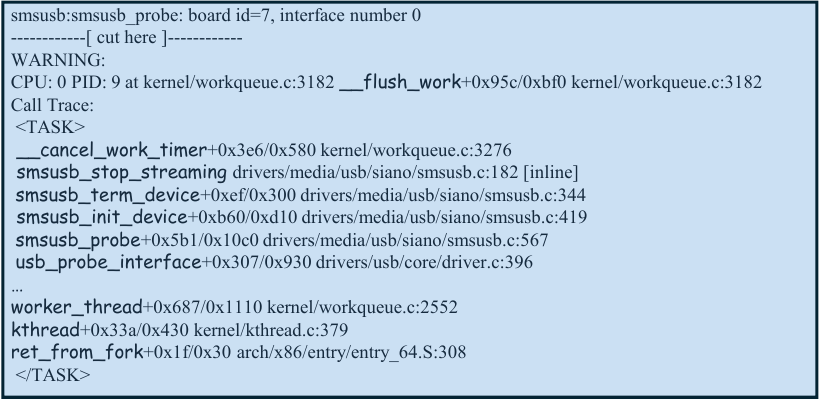}
    \caption{Crash report of the kernel crash example discussed in Appendix~\ref{app:examples}.}
    \label{fig:complete_example_crash_report}
\end{figure}

\begin{figure}[H]
    \centering
    \includegraphics[width=\linewidth]{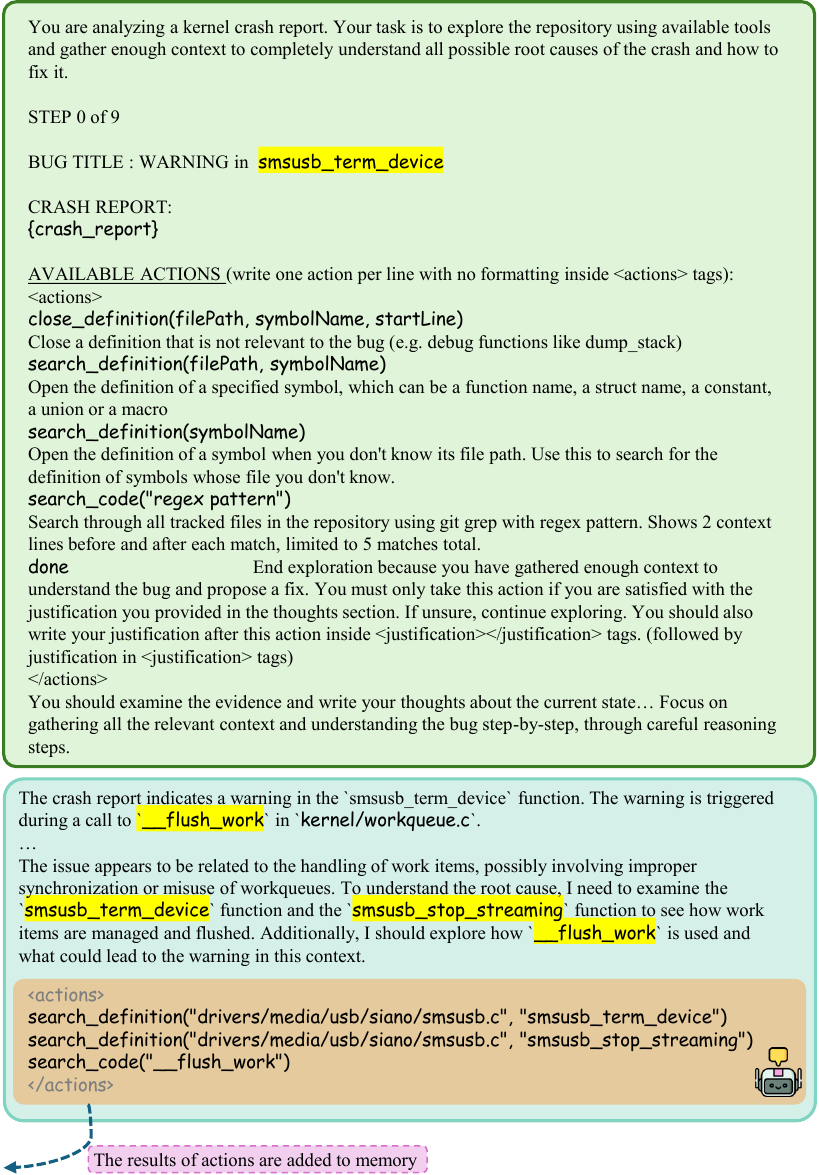}
    \caption{\cresearcher{} trajectory for the crash described in Figure~\ref{fig:complete_example_crash_report} and discussed in Appendix~\ref{app:examples}: Step 0 of 9} 
    \label{fig:complete_example_crash_report_step0}
\end{figure}
\begin{figure}[H]
    \centering
    \includegraphics[width=\linewidth]{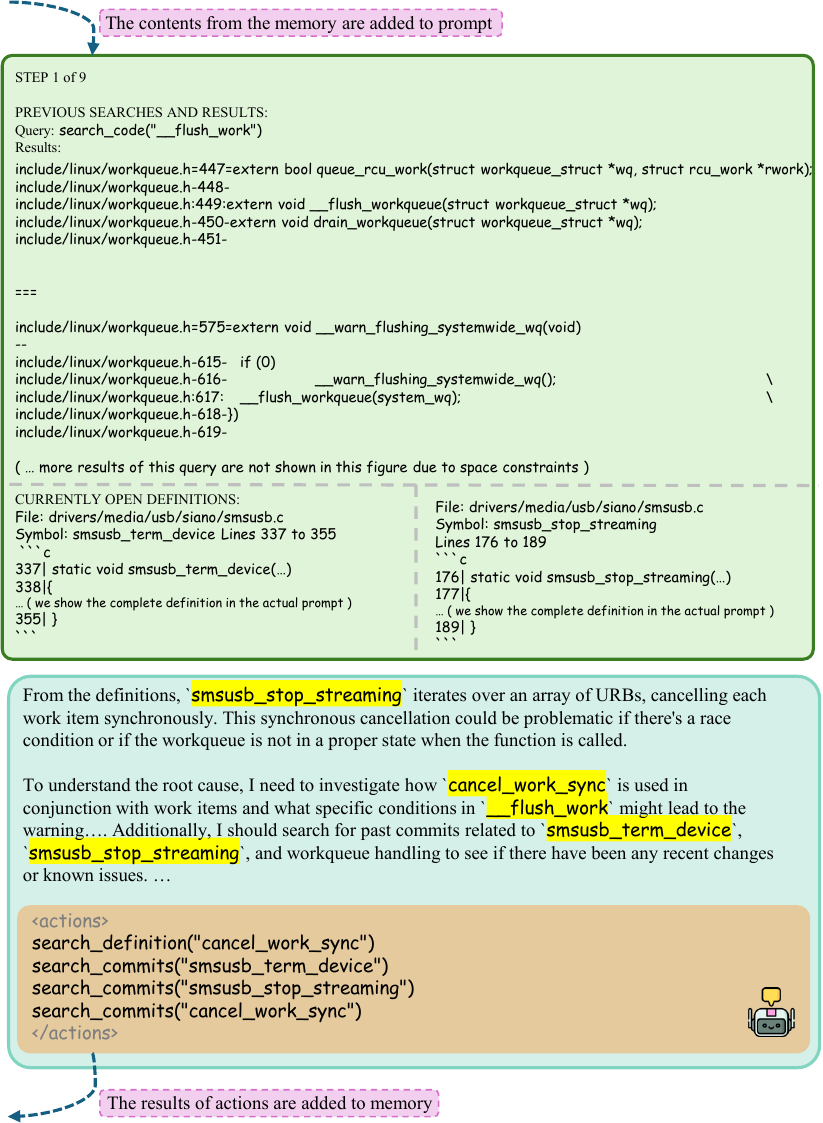}
    \caption{\cresearcher{} trajectory for the crash described in Figure~\ref{fig:complete_example_crash_report} and discussed in Appendix~\ref{app:examples}: Step 1 of 9} 
    \label{fig:complete_example_crash_report_step1}
\end{figure}

\begin{figure}[H]
    \centering
    \includegraphics[width=\linewidth]{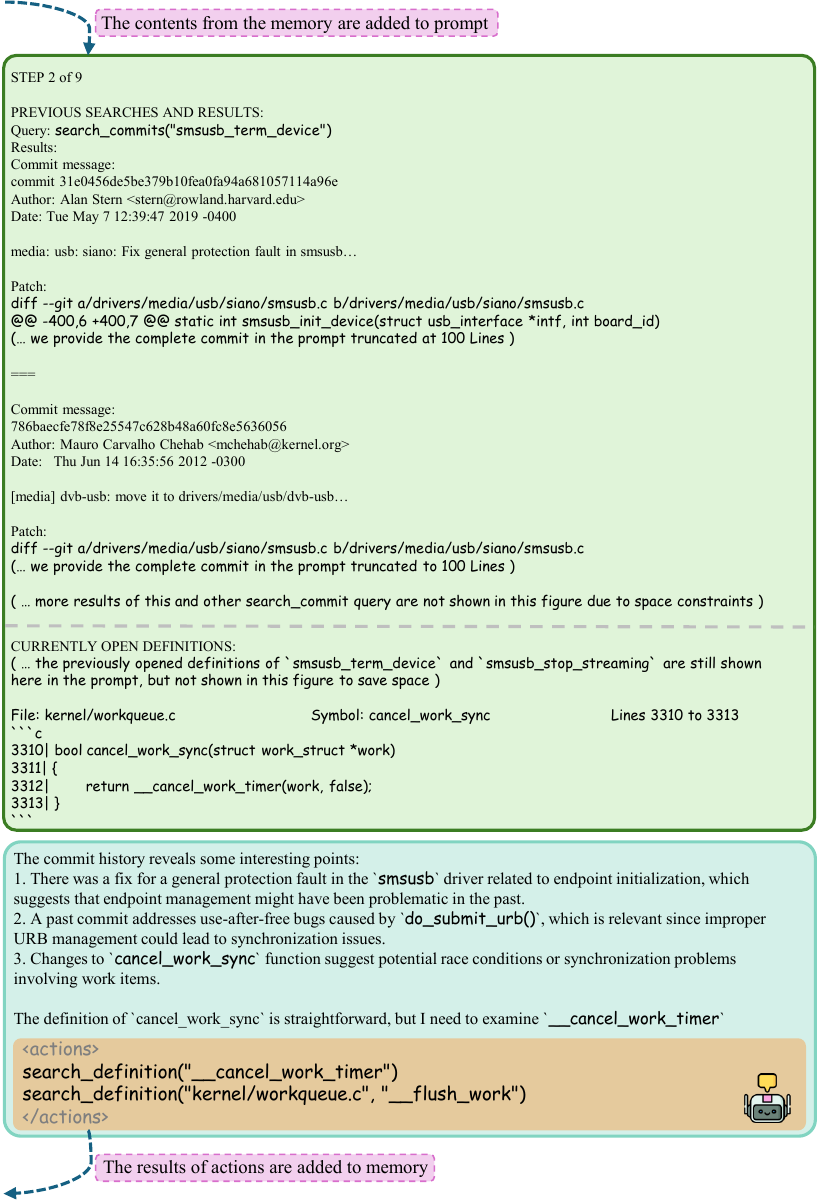}
    \caption{\cresearcher{} trajectory for the crash described in Figure~\ref{fig:complete_example_crash_report} and discussed in Appendix~\ref{app:examples}: Step 2 of 9} 
    \label{fig:complete_example_crash_report_step2}
\end{figure}

\begin{figure}[H]
    \centering
    \includegraphics[width=\linewidth]{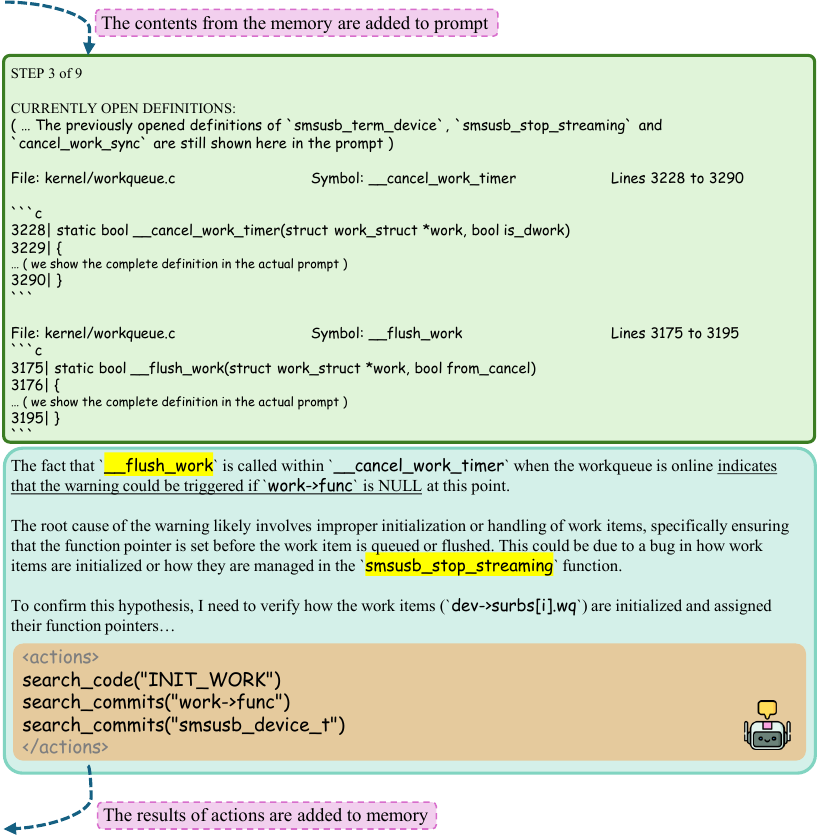}
    \caption{\cresearcher{} trajectory for the crash described in Figure~\ref{fig:complete_example_crash_report} and discussed in Appendix~\ref{app:examples}: Step 3 of 9} 
    \label{fig:complete_example_crash_report_step3}
\end{figure}

\begin{figure}[H]
    \centering
    \includegraphics[width=\linewidth]{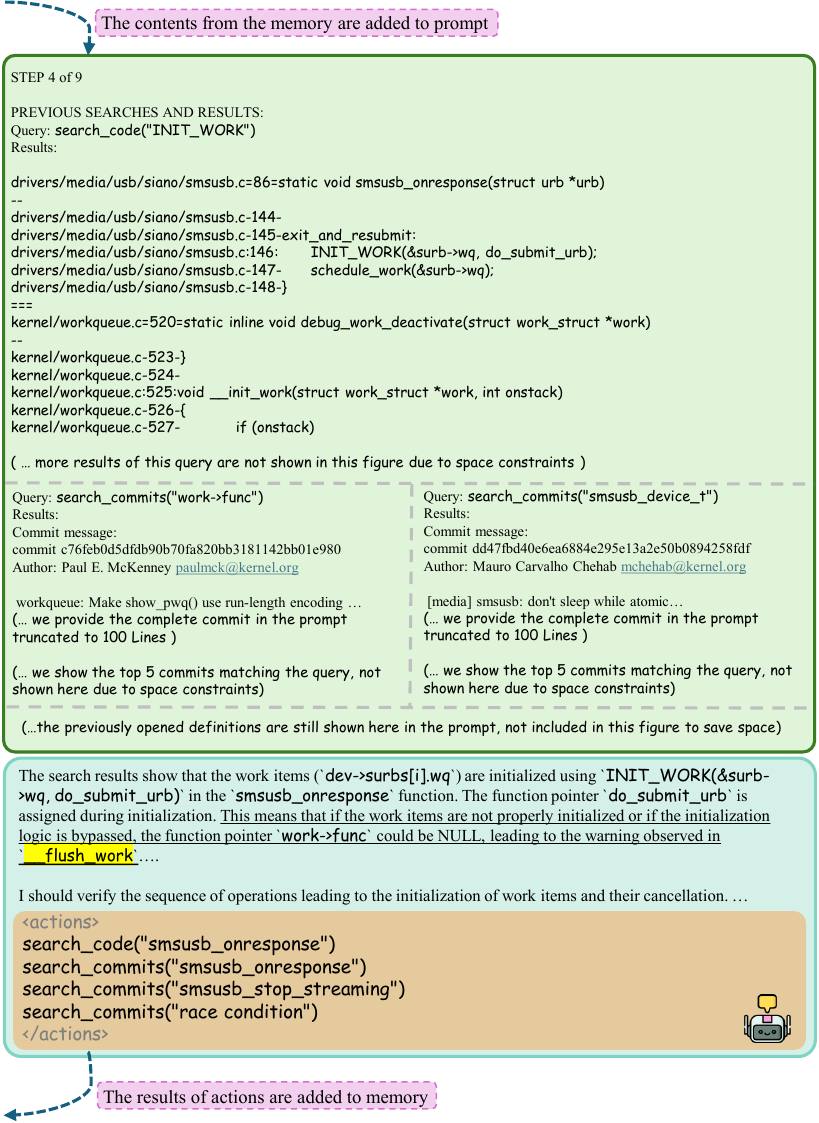}
    \caption{\cresearcher{} trajectory for the crash described in Figure~\ref{fig:complete_example_crash_report} and discussed in Appendix~\ref{app:examples}: Step 4 of 9} 
    \label{fig:complete_example_crash_report_step4}
\end{figure}

\begin{figure}[H]
    \centering
    \includegraphics[width=\linewidth]{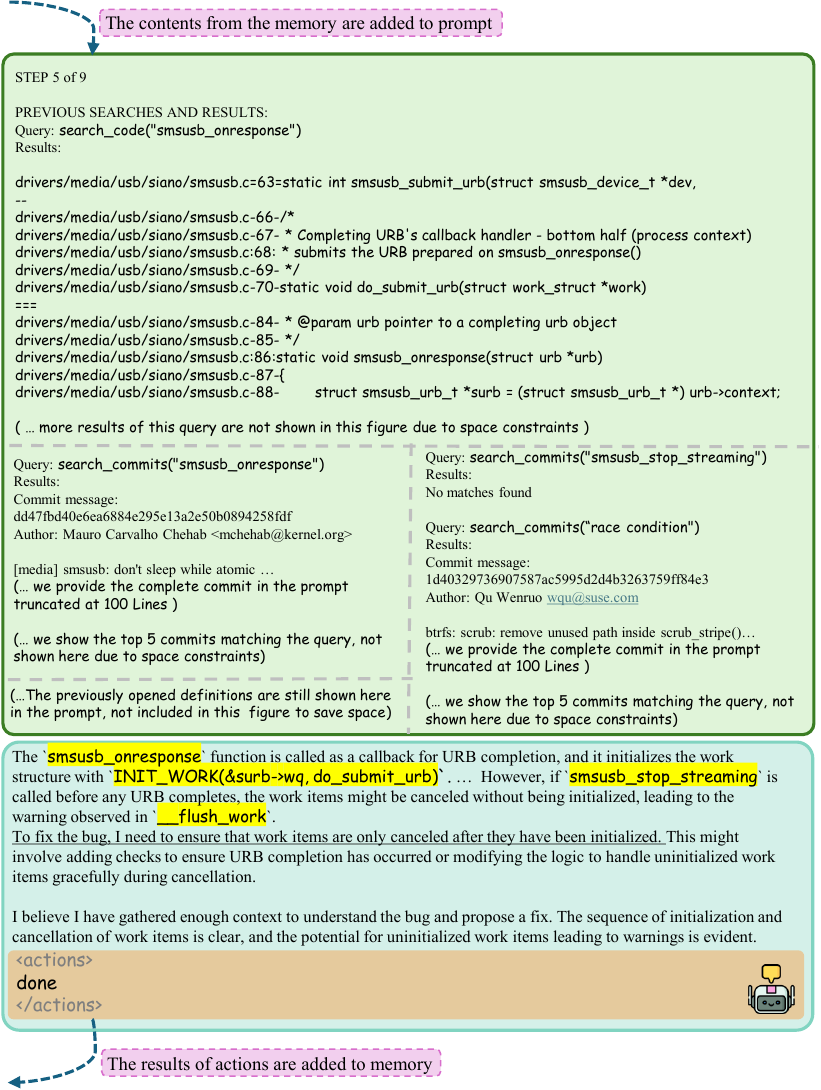}
    \caption{\cresearcher{} trajectory for the crash described in Figure~\ref{fig:complete_example_crash_report} and discussed in Appendix~\ref{app:examples}: Step 5 of 9 \analysisphase{} phase} 
    \label{fig:complete_example_crash_report_step5}
\end{figure}

%% file: appendix/long-context.tex
\section{Additional related work: long context reasoning}
\label{app:long-context}

Support for increasing context length sizes in LLMs has been an active area of research~\citep{google-long-context,longcoder},
opening up the possibility of feeding the entire repository into an LLM's context and generating a patch. But there are a few complications.
First, note that the Linux kernel has over $75$K files and $28$ Million lines of code.
In contrast, state of the art models today (e.g., Gemini 2.5 Pro) support at most $2$M tokens in the context window~\citep{google-long-context-docs,google-long-context-docs-gemini2.5}, roughly corresponding to around $100$K lines of code~\citep{google-long-context-docs}.
Second, long-context models do not robustly make use of the information in context. They often get ``lost in the middle''~\citep{lost-in-the-middle}, performing highest when relevant information occurs at the beginning or end of the input context, and significantly worse when they must access relevant information in the middle of long contexts.
\Citet{longicl} found that long-context LLMs struggle with processing long, context-rich sequences and reasoning over multiple pieces of information (which is important for any automated software development task).

%% file: appendix/exp-details.tex
\section{Experimental setup: additional details} \label{app:details}

\paragraph{Dataset details}
We use the \kbenchsyz{} dataset containing $279$ instances from \citet{kgym}. The dataset is publicly available at \url{https://github.com/Alex-Mathai-98/kGym-Kernel-Playground} and is under an MIT License.
We validated the $279$ instances (i.e., the reproducers and the ground-truth fixes), and ruled out $9$ instances for which we could not run the kernel at the parent commit, $27$ for which the kernel at the parent commit did not crash, and $43$ where the kernel still crashed after applying the fix. So, for our experiments, we use the remaining \textbf{200 instances} that we successfully validated. For reproducibility, we use the crash reports generated during our validation instead of the crash reports originally present in \kbenchsyz{}.
The subset of $200$ instances that we were able to reproduce (containing the bug ids and the crash reports from our reproduction run) is available in the file \texttt{data/kBenchSyz/200\_subset.json} in the supplementary material.

\paragraph{Sampling details}
In the \synthesisphase{} phase, we ask the agent to generate a hypothesis and patch in the following format.
It has to write the hypothesis inside \texttt{<hypothesis>} tags and the patch inside \texttt{<patch>} tags.
The content inside the \texttt{<patch>} tags is a list of \texttt{<symbol>} tags covering all the symbols whose definitions the agent wants to change in its patch.
With each tag, the agent has to provide \texttt{file, name and start line} attributes and inside each tag, it has to rewrite the complete definition of the symbol (after making the desired changes).
We use successively higher temperatures ($0$, $0.3$, $0.6$) until the agent gives a correctly formatted patch.
For o1, since its API does not support a temperature parameter, we sample the desired number of patches by setting the \texttt{n} parameter (number of completions) in the OpenAI Chat Completions API. 

\paragraph{\ffmpeg{} details}
Please refer to Appendix~\ref{app:ffmpeg} for complete details about our experiments on the \ffmpeg{} dataset.

\paragraph{Crash reproduction setup}
Our setup for building the Linux kernel and running it on reproducer files is built on top of the \kgym{} platform (MIT Licensed, publicly available at \url{https://github.com/Alex-Mathai-98/kGym-Kernel-Gym}) \citep{kgym} and has a couple of major modifications. First, while \kgym{} runs only on the Google Cloud platform, our setup can run locally on any machine and uses cloud storage for preserving compiled kernels, crash reports, etc. Second, we use \texttt{ccache} \citep{ccache} for caching build files generated during kernel compilation and our own logic for caching \texttt{git checkout}s.

\kgym{} has a distributed setup featuring five workers - \kbuilder{}, \kreproducer{}, \kscheduler{}, \kdashboard{} and \kmq{}. (1)~\kbuilder{} takes as input a source commit, a kernel config, and (optionally) a patch. It checks out the kernel at the source commit, applies the patch, compiles the kernel and uploads the build artifacts (kernel image, vmlinux binary, etc.) to cloud storage. (2)~\kreproducer{} takes as input the build artifacts and a reproducer file and runs the kernel on the reproducer while monitoring for crashes. To handle non-deterministic bugs, we launch $4$ VMs in parallel, each of which runs the reproducer. Each VM further runs multiple processes where system calls can execute in parallel so concurrency bugs can also be reproduced. If any of these VMs crash within $10$ minutes or if \kreproducer{} loses connection to the VMs, we say that the kernel crashes on the reproducer. It then uploads the crash reports to cloud storage. (3)~\kscheduler{} serves an API where we can send reproduction jobs with the source commit, config, reproducer and (optionally) patch. It communicates with \kbuilder{} and \kreproducer{} through the message queue \kmq{} and orchestrates the overall flow of build with \kbuilder{} followed by reproduction with \kreproducer{}. (4)~Finally, \kdashboard{} displays each job's logs and results in a web UI.

\paragraph{KUnit testing details}
For each of the $116$ crashes resolved by \cresearcher{} (GPT-4o+o1, P@$5$, $15$ \maxcalls{}), we selected one crash-resolving patch and ran KUnit~\citep{kunit} tests on it.
For $13$ crashes, KUnit was not present in the kernel source code version on which the crash was reported.
For another $15$ crashes, KUnit was present but did not support the CLI arguments for running tests in QEMU (required as our host kernel is different from the test kernel).
From the remaining $88$ crashes, all the KUnit tests had a status of either \texttt{PASS} or \texttt{SKIP} (some hardware-specific tests are skipped depending on the machine requirements).
On average, only $\sim 13$ tests were skipped for a patch while $\sim 210$ tests were passed.
Due to a bug in the KUnit test-suite at the parent of the fix commit for $4$ crashes, the \texttt{example\_skip\_test} and \texttt{example\_mark\_skipped\_test} tests, which should have been skipped, were run.
Similarly, for $2$ crashes, due to a bug in KUnit\footnote{See \url{https://groups.google.com/g/kunit-dev/c/ahWFBJsIA2U} and \url{https://www.spinics.net/lists/kernel/msg5128827.html}.}, \texttt{hw\_breakpoint} tests, that should have been skipped, were run.
We ignore the results of these tests as they are not relevant to the correctness of the patches being tested.

\paragraph{Compute resources}
We setup $10$ replicas of the distributed setup (containing $5$ workers) described above.
Each machine was equipped with an AMD EPYC 7V13 Processor running at $2.50$ GHz, had $24$ cores and $220$ GB RAM.
For one evaluation run on our dataset of $200$ instances for any tool in the P@$5$ setting (i.e., for evaluating $1000$ patches on whether they prevent a crash or not), we divided the instances among the $10$ replicas, and the overall time ranged from $10$ to $15$ hours.

\paragraph{\cresearcher{} Prompts}
The system and user prompts for the \analysisphase{} and \synthesisphase{} phases (for both filtering and patch generation) are provided in the \texttt{prompts} folder of our supplementary material. They use a \texttt{prompt\_preamble} and \texttt{prompt\_analysis\_examples} that vary based on the codebase (e.g., Linux kernel or FFmpeg). Those can be found in the files \texttt{config/kBenchSyz.yaml} and \texttt{config/ffmpeg.yaml} respectively in our supplementary material.

\paragraph{SWE-agent details}
We use SWE-agent~\citep{yang2024sweagentagentcomputerinterfacesenable} as one of our baselines. The codebase is publicly available at \url{https://github.com/SWE-agent/SWE-agent/tree/main} and is under the MIT License. We use version $1.0.1$ of SWE-agent, and add a Linux kernel-specific example trajectory and background about the Linux kernel to its prompts.
We provide the complete configuration file (including all prompts and the example trajectory) in the \texttt{SWE-agent} folder of our supplementary material.

\paragraph{Agentless details}
We adopt Agentless~\citep{xia2024agentlessdemystifyingllmbasedsoftware} as one of our baselines. The codebase is publicly available at \url{https://github.com/OpenAutoCoder/Agentless} and is under MIT License. The default Agentless pipeline consists of the following steps: (i) retrieval of two sets of relevant files (via an LLM and via embeddings), (ii) identification of candidate edit locations, (iii) generation of multiple patch candidates, and (iv) ranking of patches using test executions.
In our implementation, we fix the temperature at $0.1$ for all stages of the pipeline and sample $5$ candidate patches in the third stage of patch generation.
For scalability reasons, we omit the embedding-based retrieval step and ranking is immaterial in our setting since we use Pass@$5$.
Although the original implementation could not be directly reused, since it is designed specifically for Python codebases, we reimplemented the stages of the pipeline for our usecase, incorporating kernel-specific prompts at each step.

\textbf{Rationale for omitting embedding-based retrieval}
To estimate the computational overhead, we randomly sampled $1,000$ \texttt{.c} and \texttt{.h} files from the Linux kernel and computed the average number of chunks per file. Extrapolating to the entire codebase, we estimate approximately $781,889$ chunks. With the \texttt{text-embedding-3-small} model (used by Agentless), each embedding call requires on average $\sim 0.5$ seconds. This implies a total embedding time of roughly $108$ hours for a single kernel snapshot, making this step infeasible in our setting. 

All prompts used for the Agentless stages in our experiments are included in the \texttt{Agentless} folder of the supplementary material.

%% file: appendix/commit-importance.tex
\section{Example showing the importance of causal analysis over historical commits} \label{app:commit-importance}

\begin{figure}[H]
    \centering
    \includegraphics[width=\linewidth]{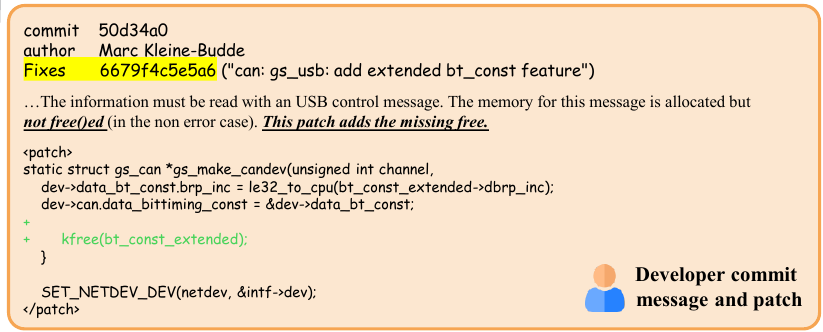}
    \caption{Developer commit message and patch.}
    \label{fig:code_a}
\end{figure}

\begin{figure}[H]
    \centering
    \includegraphics[width=\linewidth]{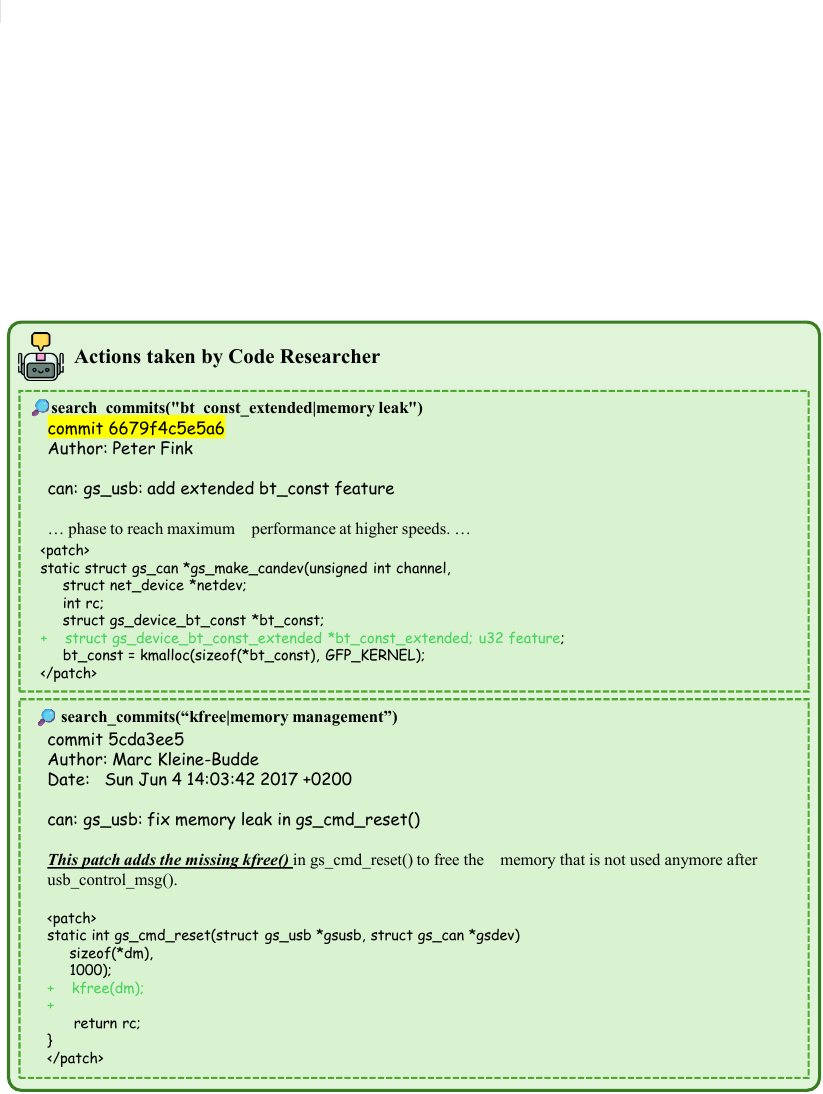}
    \caption{\cresearcher{} actions (\texttt{search\_commits} in green box).}
    \label{fig:code_b}
\end{figure}

\begin{figure}[H]
    \centering
    \includegraphics[width=\linewidth]{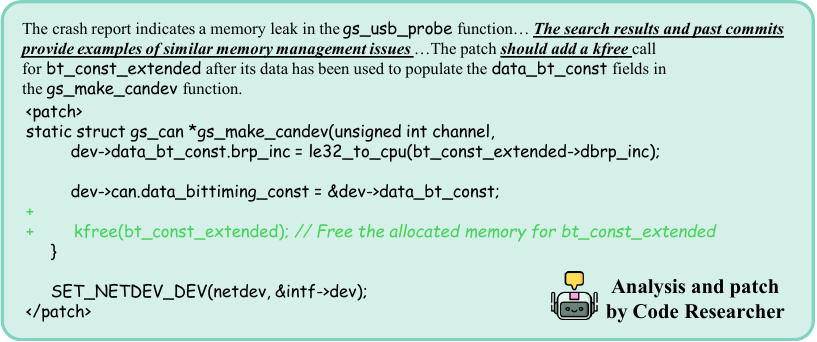}
    \caption{\cresearcher{} patch and analysis.}
    \label{fig:code_c}
\end{figure}

\noindent\textbf{Figures~\ref{fig:code_a}--\ref{fig:code_c}:}
Illustration of \cresearcher{} analyzing and repairing a real-world memory leak bug\footnote{Bug in Syzbot dashboard: \url{https://syzkaller.appspot.com/bug?id=92a742e993c8b9e769f8502a0497c88c0afa78af}.} from the \kbenchsyz{} dataset~(the complete trajectory of \cresearcher{} is truncated and only the relevant parts are shown due to space constraints).
Figure~\ref{fig:code_a} shows the developer’s original commit, including the fix and a "Fixes:" tag that references the buggy commit where the issue originated: commit \colorbox{highlightyellow}{\texttt{6679f4c5e5a6}}—highlighted in yellow. This section is shown in the \textcolor{black}{\fcolorbox{black}{orangebox}{orange box}}. The developer's fix is available at the following link\footnote{Developer’s fix commit: \url{https://git.kernel.org/pub/scm/linux/kernel/git/torvalds/linux.git/commit/?id=50d34a0d151dc7abbdbec781bd7f09f2b3cbf01a}.}. Figure~\ref{fig:code_b} displays a subset of actions taken by \cresearcher{} in the \analysisphase{} phase, specifically several \texttt{search\_commits} steps (\textcolor{black}{\fcolorbox{black}{greenbox}{green box}}), which reveal how the agent retrieves both the buggy commit and other related commits that involve memory management. These retrieved historical commits provide guidance in the \synthesisphase{} phase. Figure~\ref{fig:code_c} presents the patch and analysis synthesized by \cresearcher{} (\textcolor{black}{\fcolorbox{black}{bluebox}{blue box}}) during the \synthesisphase{} phase, which correctly identifies the missing deallocation and inserts the appropriate \texttt{kfree(bt\_const\_extended)} call. Notably, \cresearcher{} successfully navigates to the same buggy commit identified by the developer’s "Fixes:" tag, demonstrating its ability to infer causality and leverage prior commits to localize and fix bugs.

%% file: appendix/llm-as-judge.tex
\section{LLM-as-Judge evaluation of overlap between developer commit and tool-gathered context}
\label{app:llm-as-judge}
We use LLM-as-judge to analyze the context gathered by \cresearcher{} and SWE-agent to determine the overlap of context in their trajectory with the context mentioned by the developer in the ground-truth fix commit message. We first identify code symbols mentioned in the commit message for a given bug $b$, which we denote as $s^*_{b}$. Then for each candidate patch $i$, we find the overlap of $s^*_b$ with the symbols whose definitions are seen in its trajectory. We denote this overlap by $s_{b, i}$. We define symbol ratio $SR$ for each candidate patch as $$SR_{b,i} = \frac{ |s_{b,i}|}{|s^*_{b}|} .$$
We consider patch $i$ to have overlapping symbol context with the developer commit if $SR_{b,i} \ge 0.33$. We label all candidate patches with this criterion. As mentioned in Section~\ref{sec:rq3} (2), we find that SWE-agent has $54.18\%$ overlapping symbol context patches, while \cresearcher{} has $63.7\%$ overlapping symbol context patches. This indicates that \cresearcher{} is more effective at identifying relevant context. 

Additionally, we also measure the impact of finding relevant context on the crash resolution rate (CRR) as:
$$P(\text{patch resolves crash} \;|\; \text{overlapping symbol context}) = 0.309, $$
$$P(\text{patch resolves crash} \;|\; \text{non-overlapping symbol context}) = 0.116.$$
This suggests that patches with overlapping symbol context have a significantly higher probability of resolving crashes than patches without. 

In addition to symbols, we also identify commit IDs mentioned in the commit message for a given bug $b$ which we denote as $c^*_b$. Then for each candidate patch $i$, we find the overlap of $c^*_b$ with the commits retrieved in its trajectory, denoted as $c_{b,i}$.
We note that $c^*_b$ is typically a small number, with a maximum value of $3$ in our dataset of $200$ bugs.
Therefore, instead of a ratio, we label patch $i$ to have overlapping commit context when all the commits in $c^*_b$ are present in $c_{b,i}$ (i.e., $\frac{|c_{b,i}|}{|c^*_b|} = 1$).
We find that $30.8\%$ of patches produced by \cresearcher{} have overlapping commit context (recall that SWE-agent does not search over commit IDs). Further, we find that overlapping commit context also has a positive impact on CRR: $$P(\text{patch resolves crash} \;|\; \text{overlapping commit context}) = 0.315,$$ $$P(\text{patch resolves crash} \;|  \text{ non-overlapping commit context}) = 0.205.$$

Overall, these results bring out the utility of effective context retrieval.

%% file: appendix/qualitative_examples.tex
\section{Qualitative evaluation and examples}
\label{app:qualexamples}

\paragraph{Example A: \texttt{jfs\_dmap.c} boundary check.}
Listing~\ref{lst:jfs_semantic}\footnote{Example A in Syzkaller: \url{https://syzkaller.appspot.com/bug?id=5eb8a5d29d77f8a364cf3270bf9625eb4d4ffc52}}   compares the developer’s ground-truth patch  with
the patch generated by \cresearcher{}.  Both fixes add a
lower-bound check on \verb|bmp->db_agl2size| alongside the existing upper-bound
check; the only difference is the ordering of the two disjuncts in the \texttt{if}
condition, an immaterial variation in this case.
This illustrates the class of \emph{Accurate} patches.

\begin{code}
\captionof{listing}{Semantically equivalent patch
produced by \cresearcher{} for the \texttt{jfs\_dmap.c} crash.}
\label{lst:jfs_semantic}
\begin{minted}[linenos,breaklines,xleftmargin=1.5em,numbersep=5pt]{diff}
--- a/fs/jfs/jfs_dmap.c                         /* developer */
+++ b/fs/jfs/jfs_dmap.c
@@ -193,7 +193,8 @@ int dbMount(struct inode *ipbmap)
    bmp->db_agwidth = le32_to_cpu(dbmp_le->dn_agwidth);
    bmp->db_agstart = le32_to_cpu(dbmp_le->dn_agstart);
    bmp->db_agl2size = le32_to_cpu(dbmp_le->dn_agl2size);
-   if (bmp->db_agl2size > L2MAXL2SIZE - L2MAXAG) {
+   if (bmp->db_agl2size > L2MAXL2SIZE - L2MAXAG ||
+       bmp->db_agl2size < 0) {
        err = -EINVAL;
        goto err_release_metapage;
    }

--- a/fs/jfs/jfs_dmap.c                         /* generated */
+++ b/fs/jfs/jfs_dmap.c
@@ -193,7 +193,7 @@ int dbMount(struct inode *ipbmap)
    bmp->db_agwidth = le32_to_cpu(dbmp_le->dn_agwidth);
    bmp->db_agstart = le32_to_cpu(dbmp_le->dn_agstart);
    bmp->db_agl2size = le32_to_cpu(dbmp_le->dn_agl2size);
-   if (bmp->db_agl2size > L2MAXL2SIZE - L2MAXAG) {
+   if (bmp->db_agl2size < 0 ||  bmp->db_agl2size > L2MAXL2SIZE - L2MAXAG) {
        err = -EINVAL;
        goto err_release_metapage;
    }
\end{minted}
\end{code}

\paragraph{Example B: \texttt{hci\_h5.c} null-check addition.}
In Listing~\ref{lst:h5_overspec}\footnote{Example B in Syzkaller: \url{https://syzkaller.appspot.com/bug?id=1bc76a5779300075e3466a0e2abcd753df212779}}, both the developer and \cresearcher{} address an unsafe access to \texttt{hu->serdev->dev}, but the generated patch goes beyond the developer's fix. While the developer simply guards the power management calls with a null check, \cresearcher{} adds an \texttt{else} branch that logs an error and returns \texttt{-ENODEV}. This reflects a conservative design that prevents execution in the event of a null pointer, even though the surrounding kernel code may already guarantee that \texttt{hu->serdev} is non-null. Such overspecialization can be seen as a benign deviation: the patch is functionally correct and improves robustness, but at the risk of silently diverging from upstream assumptions. This illustrates the class of \emph{overspecialized} patches.

\begin{code}
\captionof{listing}{Overspecialized patch generated for \texttt{hci\_h5.c}.}
\label{lst:h5_overspec}
\begin{minted}[linenos,breaklines,xleftmargin=1.5em,numbersep=5pt]{diff}
--- a/drivers/bluetooth/hci_h5.c                /* developer */
+++ b/drivers/bluetooth/hci_h5.c
@@ -587,9 +587,11 @@ static int h5_recv(...)
    count -= processed;
    }

-   pm_runtime_get(&hu->serdev->dev);
-   pm_runtime_mark_last_busy(&hu->serdev->dev);
-   pm_runtime_put_autosuspend(&hu->serdev->dev);
+   if (hu->serdev) {
+       pm_runtime_get(&hu->serdev->dev);
+       pm_runtime_mark_last_busy(&hu->serdev->dev);
+       pm_runtime_put_autosuspend(&hu->serdev->dev);
+   }

    return 0;
 }

--- a/drivers/bluetooth/hci_h5.c                /* generated */
+++ b/drivers/bluetooth/hci_h5.c
@@ -587,9 +587,14 @@ static int h5_recv(...)
 		count -= processed;
 	}
    
-   pm_runtime_get(&hu->serdev->dev);
-   pm_runtime_mark_last_busy(&hu->serdev->dev);
-   pm_runtime_put_autosuspend(&hu->serdev->dev);
+   if (hu->serdev) {
+       pm_runtime_get(&hu->serdev->dev);
+       pm_runtime_mark_last_busy(&hu->serdev->dev);
+       pm_runtime_put_autosuspend(&hu->serdev->dev);
+   } else {
+       bt_dev_err(hu->hdev, "serdev is not initialized");
+       return -ENODEV;
+   }

    return 0;
 }
\end{minted}
\end{code}

\paragraph{Example C: \texttt{ns.c} RCU read lock insertion.}
In Listing~\ref{lst:ns_rcu_partial}\footnote{Example C in Syzkaller: \url{https://syzkaller.appspot.com/bug?id=07c9d71dc1a215b19c6a245c68f502bc57dbdb83}} both the developer and \cresearcher{} address the unsafe traversal of a radix tree without proper RCU synchronization. The developer applies a comprehensive fix, wrapping all relevant \verb|radix_tree_for_each_slot| iterations with \verb|rcu_read_lock()| and \verb|rcu_read_unlock()| across multiple functions. In contrast, \cresearcher{} focuses only on the \verb|ctrl_cmd_new_lookup()| function, inserting the necessary locking primitives in that scope alone. While this partial patch is not directly mergeable due to its incompleteness, it demonstrates an accurate understanding of the underlying concurrency issue and correctly applies the mitigation in the context it modifies. As such, it exemplifies the class of \emph{incomplete} patches, offering concrete insight into the nature and location of the bug, and accelerating the path toward a complete and upstreamable fix.

\begin{code}
\captionof{listing}{Developer and plausible patches for \texttt{ns.c}.}
\label{lst:ns_rcu_partial}
\begin{minted}[linenos,breaklines,xleftmargin=1.5em,numbersep=5pt]{diff}
--- a/net/qrtr/ns.c                         /* developer */
+++ b/net/qrtr/ns.c
@@ -193,12 +193,13 @@ static int announce_servers(struct sockaddr_qrtr *sq)
 	struct qrtr_server *srv;
 	struct qrtr_node *node;
 	void __rcu **slot;
-	int ret;
+	int ret = 0;
 
 	node = node_get(qrtr_ns.local_node);
 	if (!node)
 		return 0;
 
+	rcu_read_lock();
 	/* Announce the list of servers registered in this node */
 	radix_tree_for_each_slot(slot, &node->servers, &iter, 0) {
 		srv = radix_tree_deref_slot(slot);
@@ -206,11 +207,14 @@ static int announce_servers(struct sockaddr_qrtr *sq)
 		ret = service_announce_new(sq, srv);
 		if (ret < 0) {
 			pr_err("failed to announce new service\n");
-			return ret;
+			goto err_out;
 		}
 	}
 
-	return 0;
+err_out:
+	rcu_read_unlock();
+
+	return ret;
 }
 
 static struct qrtr_server *server_add(unsigned int service,
@@ -335,7 +339,7 @@ static int ctrl_cmd_bye(struct sockaddr_qrtr *from)
 	struct qrtr_node *node;
 	void __rcu **slot;
 	struct kvec iv;
-	int ret;
+	int ret = 0;
 
 	iv.iov_base = &pkt;
 	iv.iov_len = sizeof(pkt);
@@ -344,11 +348,13 @@ static int ctrl_cmd_bye(struct sockaddr_qrtr *from)
 	if (!node)
 		return 0;
 
+	rcu_read_lock();
 	/* Advertise removal of this client to all servers of remote node */
 	radix_tree_for_each_slot(slot, &node->servers, &iter, 0) {
 		srv = radix_tree_deref_slot(slot);
 		server_del(node, srv->port);
 	}
+	rcu_read_unlock();
 
 	/* Advertise the removal of this client to all local servers */
 	local_node = node_get(qrtr_ns.local_node);
@@ -359,6 +365,7 @@ static int ctrl_cmd_bye(struct sockaddr_qrtr *from)
 	pkt.cmd = cpu_to_le32(QRTR_TYPE_BYE);
 	pkt.client.node = cpu_to_le32(from->sq_node);
 
+	rcu_read_lock();
 	radix_tree_for_each_slot(slot, &local_node->servers, &iter, 0) {
 		srv = radix_tree_deref_slot(slot);
 
@@ -372,11 +379,14 @@ static int ctrl_cmd_bye(struct sockaddr_qrtr *from)
 		ret = kernel_sendmsg(qrtr_ns.sock, &msg, &iv, 1, sizeof(pkt));
 		if (ret < 0) {
 			pr_err("failed to send bye cmd\n");
-			return ret;
+			goto err_out;
 		}
 	}
 
-	return 0;
+err_out:
+	rcu_read_unlock();
+
+	return ret;
 }
 
 static int ctrl_cmd_del_client(struct sockaddr_qrtr *from,
@@ -394,7 +404,7 @@ static int ctrl_cmd_del_client(struct sockaddr_qrtr *from,
 	struct list_head *li;
 	void __rcu **slot;
 	struct kvec iv;
-	int ret;
+	int ret = 0;
 
 	iv.iov_base = &pkt;
 	iv.iov_len = sizeof(pkt);
@@ -434,6 +444,7 @@ static int ctrl_cmd_del_client(struct sockaddr_qrtr *from,
 	pkt.client.node = cpu_to_le32(node_id);
 	pkt.client.port = cpu_to_le32(port);
 
+	rcu_read_lock();
 	radix_tree_for_each_slot(slot, &local_node->servers, &iter, 0) {
 		srv = radix_tree_deref_slot(slot);
 
@@ -447,11 +458,14 @@ static int ctrl_cmd_del_client(struct sockaddr_qrtr *from,
 		ret = kernel_sendmsg(qrtr_ns.sock, &msg, &iv, 1, sizeof(pkt));
 		if (ret < 0) {
 			pr_err("failed to send del client cmd\n");
-			return ret;
+			goto err_out;
 		}
 	}
 
-	return 0;
+err_out:
+	rcu_read_unlock();
+
+	return ret;
 }
 
 static int ctrl_cmd_new_server(struct sockaddr_qrtr *from,
@@ -554,6 +568,7 @@ static int ctrl_cmd_new_lookup(struct sockaddr_qrtr *from,
 	filter.service = service;
 	filter.instance = instance;
 
+	rcu_read_lock();
 	radix_tree_for_each_slot(node_slot, &nodes, &node_iter, 0) {
 		node = radix_tree_deref_slot(node_slot);
 
@@ -568,6 +583,7 @@ static int ctrl_cmd_new_lookup(struct sockaddr_qrtr *from,
 			lookup_notify(from, srv, true);
 		}
 	}
+	rcu_read_unlock();
 
 	/* Empty notification, to indicate end of listing */
 	lookup_notify(from, NULL, true);

--- a/net/qrtr/ns.c                /* generated */
+++ b/net/qrtr/ns.c
@@ -554,6 +554,7 @@ static int ctrl_cmd_new_lookup(struct sockaddr_qrtr *from,
 	filter.service = service;
 	filter.instance = instance;

+	rcu_read_lock(); // Add RCU read lock
 	radix_tree_for_each_slot(node_slot, &nodes, &node_iter, 0) {
 		node = radix_tree_deref_slot(node_slot);

@@ -568,6 +569,7 @@ static int ctrl_cmd_new_lookup(struct sockaddr_qrtr *from,
 			lookup_notify(from, srv, true);
 		}
 	}
+	rcu_read_unlock(); // Add RCU read unlock

 	/* Empty notification, to indicate end of listing */
 	lookup_notify(from, NULL, true);
\end{minted}
\end{code}

\paragraph{Example D: \texttt{qrtr.c} port validation.}
In Listing~\ref{lst:qrtr_misfix}\footnote{Example D in Syzkaller: \url{https://syzkaller.appspot.com/bug?id=ca2299cf11b3e3d3d0f44ac479410a14eecbd326}}, the developer replaces \texttt{idr\_alloc()} with \texttt{idr\_alloc\_u32()} to avoid casting the (possibly large) \texttt{u32} port number to \texttt{int}.
By contrast, \cresearcher{} adds defensive checks that reject ports with \texttt{port < 0}, both in \texttt{qrtr\_port\_assign} and \texttt{\_\_qrtr\_bind}.
This patch resolves the crash, but rejects certain port numbers rather than handling them, so is not equivalent to the developer patch and is \emph{inaccurate}.
But the incoming value originates from \texttt{\_\_u32 sq\_port}, and special constants like \texttt{QRTR\_PORT\_CTRL} (defined as \texttt{0xfffffffeu}) are valid and widely used in the subsystem.

\begin{code}
\captionof{listing}{Developer and inaccurate patches for \texttt{qrtr.c}.}
\label{lst:qrtr_misfix}
\begin{minted}[linenos,breaklines,xleftmargin=1.5em,numbersep=5pt]{diff}
--- a/net/qrtr/qrtr.c                         /* developer */
+++ b/net/qrtr/qrtr.c
@@ -692,23 +692,25 @@ static void qrtr_port_remove(struct qrtr_sock *ipc)
  */
 static int qrtr_port_assign(struct qrtr_sock *ipc, int *port)
 {
+	u32 min_port;
 	int rc;

 	mutex_lock(&qrtr_port_lock);
 	if (!*port) {
-		rc = idr_alloc(&qrtr_ports, ipc,
-			       QRTR_MIN_EPH_SOCKET, QRTR_MAX_EPH_SOCKET + 1,
-			       GFP_ATOMIC);
-		if (rc >= 0)
-			*port = rc;
+		min_port = QRTR_MIN_EPH_SOCKET;
+		rc = idr_alloc_u32(&qrtr_ports, ipc, &min_port, QRTR_MAX_EPH_SOCKET, GFP_ATOMIC);
+		if (!rc)
+			*port = min_port;
 	} else if (*port < QRTR_MIN_EPH_SOCKET && !capable(CAP_NET_ADMIN)) {
 		rc = -EACCES;
 	} else if (*port == QRTR_PORT_CTRL) {
-		rc = idr_alloc(&qrtr_ports, ipc, 0, 1, GFP_ATOMIC);
+		min_port = 0;
+		rc = idr_alloc_u32(&qrtr_ports, ipc, &min_port, 0, GFP_ATOMIC);
 	} else {
-		rc = idr_alloc(&qrtr_ports, ipc, *port, *port + 1, GFP_ATOMIC);
-		if (rc >= 0)
-			*port = rc;
+		min_port = *port;
+		rc = idr_alloc_u32(&qrtr_ports, ipc, &min_port, *port, GFP_ATOMIC);
+		if (!rc)
+			*port = min_port;
 	}
 	mutex_unlock(&qrtr_port_lock);

--- a/net/qrtr/qrtr.c                /* generated */
+++ b/net/qrtr/qrtr.c
@@ -695,7 +695,9 @@ static int qrtr_port_assign(struct qrtr_sock *ipc, int *port)
 	int rc;

 	mutex_lock(&qrtr_port_lock);
-	if (!*port) {
+	if (*port < 0) {
+		rc = -EINVAL;
+	} else if (!*port) {
 		rc = idr_alloc(&qrtr_ports, ipc,
 			       QRTR_MIN_EPH_SOCKET, QRTR_MAX_EPH_SOCKET + 1,
 			       GFP_ATOMIC);
@@ -759,6 +761,9 @@ static int __qrtr_bind(struct socket *sock,
 		return 0;

 	port = addr->sq_port;
+	if (port < 0)
+		return -EINVAL;
+
 	rc = qrtr_port_assign(ipc, &port);
 	if (rc)
 		return rc;
\end{minted}
\end{code}

%% file: iclr2026_conference.bbl
\begin{thebibliography}{44}
\providecommand{\natexlab}[1]{#1}
\providecommand{\url}[1]{\texttt{#1}}
\expandafter\ifx\csname urlstyle\endcsname\relax
  \providecommand{\doi}[1]{doi: #1}\else
  \providecommand{\doi}{doi: \begingroup \urlstyle{rm}\Url}\fi

\bibitem[Anthropic(2025)]{anthropicdr}
Anthropic.
\newblock Claude takes research to new places, 2025.
\newblock URL \url{https://www.anthropic.com/news/research}.

\bibitem[{ccache}()]{ccache}
{ccache}.
\newblock ccache.
\newblock URL \url{https://github.com/ccache/ccache}.

\bibitem[Chen et~al.(2025)Chen, Tang, Deng, Wu, Wu, Jiang, Prasanna, Cohan, and
  Wang]{locagent}
Zhaoling Chen, Xiangru Tang, Gangda Deng, Fang Wu, Jialong Wu, Zhiwei Jiang,
  Viktor Prasanna, Arman Cohan, and Xingyao Wang.
\newblock Locagent: Graph-guided llm agents for code localization.
\newblock \emph{arXiv preprint arXiv:2503.09089}, 2025.

\bibitem[Engler et~al.(2001)Engler, Chen, Hallem, Chou, and Chelf]{deviant}
Dawson Engler, David~Yu Chen, Seth Hallem, Andy Chou, and Benjamin Chelf.
\newblock Bugs as deviant behavior: a general approach to inferring errors in
  systems code.
\newblock \emph{SIGOPS Oper. Syst. Rev.}, 35\penalty0 (5):\penalty0 57–72,
  October 2001.
\newblock ISSN 0163-5980.
\newblock \doi{10.1145/502059.502041}.
\newblock URL \url{https://doi.org/10.1145/502059.502041}.

\bibitem[FFmpeg(2025)]{ffmpeg}
FFmpeg.
\newblock Ffmpeg, 2025.
\newblock URL \url{https://ffmpeg.org/}.

\bibitem[Godbole et~al.(2024)Godbole, Monath, Kim, Rawat, McCallum, and
  Zaheer]{godbole2024analysis}
Ameya Godbole, Nicholas Monath, Seungyeon Kim, Ankit~Singh Rawat, Andrew
  McCallum, and Manzil Zaheer.
\newblock Analysis of plan-based retrieval for grounded text generation.
\newblock \emph{arXiv preprint arXiv:2408.10490}, 2024.

\bibitem[{Google}()]{ossfuzz}
{Google}.
\newblock Oss-fuzz | documentation for oss-fuzz.
\newblock URL \url{https://google.github.io/oss-fuzz/}.

\bibitem[Google(2025{\natexlab{a}})]{geminidr}
Google.
\newblock Gemini deep research, 2025{\natexlab{a}}.
\newblock URL \url{https://gemini.google/overview/deep-research}.

\bibitem[Google(2025{\natexlab{b}})]{google-long-context-docs}
Google.
\newblock Generative ai | documentation | long context.
\newblock
  \url{https://cloud.google.com/vertex-ai/generative-ai/docs/long-context},
  2025{\natexlab{b}}.
\newblock Accessed: 2025-03-13.

\bibitem[Google(2025{\natexlab{c}})]{google-long-context-docs-gemini2.5}
Google.
\newblock Generative ai | documentation | gemini 2.5 pro.
\newblock
  \url{https://cloud.google.com/vertex-ai/generative-ai/docs/models/gemini/2-5-pro},
  2025{\natexlab{c}}.
\newblock Accessed: 2025-05-16.

\bibitem[{Google}(2025{\natexlab{a}})]{kasan}
{Google}.
\newblock Kasan, 2025{\natexlab{a}}.
\newblock URL
  \url{https://github.com/google/kernel-sanitizers/blob/master/KASAN.md}.

\bibitem[{Google}(2025{\natexlab{b}})]{syzkaller}
{Google}.
\newblock Syzkaller, 2025{\natexlab{b}}.
\newblock URL \url{https://github.com/google/syzkaller/}.

\bibitem[Gottweis et~al.(2025)Gottweis, Weng, Daryin, Tu, Palepu, Sirkovic,
  Myaskovsky, Weissenberger, Rong, Tanno, et~al.]{gottweis2025towards}
Juraj Gottweis, Wei-Hung Weng, Alexander Daryin, Tao Tu, Anil Palepu, Petar
  Sirkovic, Artiom Myaskovsky, Felix Weissenberger, Keran Rong, Ryutaro Tanno,
  et~al.
\newblock Towards an ai co-scientist.
\newblock \emph{arXiv preprint arXiv:2502.18864}, 2025.

\bibitem[Guo et~al.(2023)Guo, Xu, Duan, Yin, and McAuley]{longcoder}
Daya Guo, Canwen Xu, Nan Duan, Jian Yin, and Julian McAuley.
\newblock Longcoder: A long-range pre-trained language model for code
  completion.
\newblock In \emph{International Conference on Machine Learning}, pp.\
  12098--12107. PMLR, 2023.

\bibitem[Jain et~al.(2024)Jain, Han, Gu, Li, Yan, Zhang, Wang, Solar-Lezama,
  Sen, and Stoica]{jain2024livecodebench}
Naman Jain, King Han, Alex Gu, Wen-Ding Li, Fanjia Yan, Tianjun Zhang, Sida
  Wang, Armando Solar-Lezama, Koushik Sen, and Ion Stoica.
\newblock Livecodebench: Holistic and contamination free evaluation of large
  language models for code.
\newblock \emph{arXiv preprint arXiv:2403.07974}, 2024.

\bibitem[Jain et~al.(2025)Jain, Singh, Shetty, Zhang, Zheng, Sen, and
  Stoica]{regym}
Naman Jain, Jaskirat Singh, Manish Shetty, Tianjun Zhang, Liang Zheng, Koushik
  Sen, and Ion Stoica.
\newblock R2e-gym: Procedural environment generation and hybrid verifiers for
  scaling open-weights {SWE} agents.
\newblock In \emph{Second Conference on Language Modeling}, 2025.
\newblock URL \url{https://openreview.net/forum?id=7evvwwdo3z}.

\bibitem[Jimenez et~al.(2024)Jimenez, Yang, Wettig, Yao, Pei, Press, and
  Narasimhan]{jimenez2024swebench}
Carlos~E Jimenez, John Yang, Alexander Wettig, Shunyu Yao, Kexin Pei, Ofir
  Press, and Karthik~R Narasimhan.
\newblock {SWE}-bench: Can language models resolve real-world github issues?
\newblock In \emph{The Twelfth International Conference on Learning
  Representations}, 2024.
\newblock URL \url{https://openreview.net/forum?id=VTF8yNQM66}.

\bibitem[Li et~al.(2024)Li, Zhang, Do, Yue, and Chen]{longicl}
Tianle Li, Ge~Zhang, Quy~Duc Do, Xiang Yue, and Wenhu Chen.
\newblock Long-context llms struggle with long in-context learning.
\newblock \emph{arXiv preprint arXiv:2404.02060}, 2024.

\bibitem[Li et~al.(2025)Li, Jin, Dong, Qian, Zhu, Wu, Wen, and
  Dou]{li2025webthinker}
Xiaoxi Li, Jiajie Jin, Guanting Dong, Hongjin Qian, Yutao Zhu, Yongkang Wu,
  Ji-Rong Wen, and Zhicheng Dou.
\newblock Webthinker: Empowering large reasoning models with deep research
  capability.
\newblock \emph{arXiv preprint arXiv:2504.21776}, 2025.

\bibitem[{Linux}(2025)]{kunit}
{Linux}.
\newblock Kunit - linux kernel unit testing, 2025.
\newblock URL
  \url{https://www.kernel.org/doc/html/latest/dev-tools/kunit/index.html}.

\bibitem[Liu et~al.(2024)Liu, Lin, Hewitt, Paranjape, Bevilacqua, Petroni, and
  Liang]{lost-in-the-middle}
Nelson~F Liu, Kevin Lin, John Hewitt, Ashwin Paranjape, Michele Bevilacqua,
  Fabio Petroni, and Percy Liang.
\newblock Lost in the middle: How language models use long contexts.
\newblock \emph{Transactions of the Association for Computational Linguistics},
  12:\penalty0 157--173, 2024.

\bibitem[Ma et~al.(2025)Ma, Yang, Cao, Li, Huang, and Li]{lingma}
Yingwei Ma, Qingping Yang, Rongyu Cao, Binhua Li, Fei Huang, and Yongbin Li.
\newblock Alibaba lingmaagent: Improving automated issue resolution via
  comprehensive repository exploration.
\newblock In \emph{Proceedings of the 33rd ACM International Conference on the
  Foundations of Software Engineering}, pp.\  238--249, 2025.

\bibitem[Mathai et~al.(2024)Mathai, Huang, Maniatis, Nogikh, Ivan{\v{c}}i{\'c},
  Yang, and Ray]{kgym}
Alex Mathai, Chenxi Huang, Petros Maniatis, Aleksandr Nogikh, Franjo
  Ivan{\v{c}}i{\'c}, Junfeng Yang, and Baishakhi Ray.
\newblock Kgym: A platform and dataset to benchmark large language models on
  linux kernel crash resolution.
\newblock \emph{Advances in Neural Information Processing Systems},
  37:\penalty0 78053--78078, 2024.

\bibitem[Mathai et~al.(2025)Mathai, Huang, Ma, Kim, Mitchell, Nogikh, Maniatis,
  Ivan{\v{c}}i{\'c}, Yang, and Ray]{crashfixer}
Alex Mathai, Chenxi Huang, Suwei Ma, Jihwan Kim, Hailie Mitchell, Aleksandr
  Nogikh, Petros Maniatis, Franjo Ivan{\v{c}}i{\'c}, Junfeng Yang, and
  Baishakhi Ray.
\newblock Crashfixer: A crash resolution agent for the linux kernel.
\newblock \emph{arXiv preprint arXiv:2504.20412}, 2025.

\bibitem[Microsoft(2025)]{microsoftdr}
Microsoft.
\newblock Introducing researcher and analyst in microsoft 365 copilot, 2025.
\newblock URL
  \url{https://www.microsoft.com/en-us/microsoft-365/blog/2025/03/25/introducing-researcher-and-analyst-in-microsoft-365-copilot/}.

\bibitem[Nielson et~al.(2015)Nielson, Nielson, and
  Hankin]{nielson2015principles}
Flemming Nielson, Hanne~R Nielson, and Chris Hankin.
\newblock \emph{Principles of program analysis}.
\newblock springer, 2015.

\bibitem[{OpenAI}(2024{\natexlab{a}})]{gpt4o}
{OpenAI}.
\newblock Introducing openai o1-preview, 2024{\natexlab{a}}.
\newblock URL \url{https://openai.com/index/hello-gpt-4o/}.

\bibitem[{OpenAI}(2024{\natexlab{b}})]{o1}
{OpenAI}.
\newblock Introducing openai o1-preview, 2024{\natexlab{b}}.
\newblock URL \url{https://openai.com/index/introducing-openai-o1-preview/}.

\bibitem[OpenAI(2025{\natexlab{a}})]{OpenAIDRGH}
OpenAI.
\newblock Openai deep research integration with github, 2025{\natexlab{a}}.
\newblock URL
  \url{https://help.openai.com/en/articles/11145903-connecting-github-to-chatgpt-deep-research}.

\bibitem[OpenAI(2025{\natexlab{b}})]{openaidr}
OpenAI.
\newblock Introducing deep research, 2025{\natexlab{b}}.
\newblock URL \url{https://openai.com/index/introducing-deep-research/}.

\bibitem[Ouyang et~al.(2025)Ouyang, Yu, Ma, Xiao, Zhang, Jia, Han, Zhang, and
  Yu]{repograph}
Siru Ouyang, Wenhao Yu, Kaixin Ma, Zilin Xiao, Zhihan Zhang, Mengzhao Jia,
  Jiawei Han, Hongming Zhang, and Dong Yu.
\newblock Repograph: Enhancing {AI} software engineering with repository-level
  code graph.
\newblock In \emph{The Thirteenth International Conference on Learning
  Representations}, 2025.
\newblock URL \url{https://openreview.net/forum?id=dw9VUsSHGB}.

\bibitem[Perplexity(2025)]{perplexitydr}
Perplexity.
\newblock Introducing perplexity deep research, 2025.
\newblock URL
  \url{https://www.perplexity.ai/de/hub/blog/introducing-perplexity-deep-research}.

\bibitem[Shao et~al.(2024)Shao, Jiang, Kanell, Xu, Khattab, and
  Lam]{shao2024assisting}
Yijia Shao, Yucheng Jiang, Theodore~A Kanell, Peter Xu, Omar Khattab, and
  Monica~S Lam.
\newblock Assisting in writing wikipedia-like articles from scratch with large
  language models.
\newblock \emph{arXiv preprint arXiv:2402.14207}, 2024.

\bibitem[Team et~al.(2024)Team, Georgiev, Lei, Burnell, Bai, Gulati, Tanzer,
  Vincent, Pan, Wang, et~al.]{google-long-context}
Gemini Team, Petko Georgiev, Ving~Ian Lei, Ryan Burnell, Libin Bai, Anmol
  Gulati, Garrett Tanzer, Damien Vincent, Zhufeng Pan, Shibo Wang, et~al.
\newblock Gemini 1.5: Unlocking multimodal understanding across millions of
  tokens of context.
\newblock \emph{arXiv preprint arXiv:2403.05530}, 2024.

\bibitem[Torvalds(1991)]{linux}
Linus Torvalds.
\newblock Linux, 1991.
\newblock URL \url{https://github.com/torvalds/linux}.

\bibitem[{universal-ctags}()]{ctags}
{universal-ctags}.
\newblock ctags.
\newblock URL \url{https://github.com/universal-ctags/ctags}.

\bibitem[Wadhwa et~al.(2024)Wadhwa, Sonwane, Arora, Mehrotra, Utpala, Bairi,
  Kanade, and Natarajan]{wadhwa2024masai}
Nalin Wadhwa, Atharv Sonwane, Daman Arora, Abhav Mehrotra, Saiteja Utpala,
  Ramakrishna~B Bairi, Aditya Kanade, and Nagarajan Natarajan.
\newblock {MASAI}: Modular architecture for software-engineering {AI} agents.
\newblock In \emph{NeurIPS 2024 Workshop on Open-World Agents}, 2024.
\newblock URL \url{https://openreview.net/forum?id=NSINt8lLYB}.

\bibitem[Wang et~al.(2025)Wang, Li, Song, Xu, Tang, Zhuge, Pan, Song, Li,
  Singh, Tran, Li, Ma, Zheng, Qian, Shao, Muennighoff, Zhang, Hui, Lin,
  Brennan, Peng, Ji, and Neubig]{openhands}
Xingyao Wang, Boxuan Li, Yufan Song, Frank~F. Xu, Xiangru Tang, Mingchen Zhuge,
  Jiayi Pan, Yueqi Song, Bowen Li, Jaskirat Singh, Hoang~H. Tran, Fuqiang Li,
  Ren Ma, Mingzhang Zheng, Bill Qian, Yanjun Shao, Niklas Muennighoff, Yizhe
  Zhang, Binyuan Hui, Junyang Lin, Robert Brennan, Hao Peng, Heng Ji, and
  Graham Neubig.
\newblock Openhands: An open platform for {AI} software developers as
  generalist agents.
\newblock In \emph{The Thirteenth International Conference on Learning
  Representations}, 2025.
\newblock URL \url{https://openreview.net/forum?id=OJd3ayDDoF}.

\bibitem[Wu et~al.(2025{\natexlab{a}})Wu, Zhu, and Liu]{wu2025agentic}
Junde Wu, Jiayuan Zhu, and Yuyuan Liu.
\newblock Agentic reasoning: Reasoning llms with tools for the deep research.
\newblock \emph{arXiv preprint arXiv:2502.04644}, 2025{\natexlab{a}}.

\bibitem[Wu et~al.(2025{\natexlab{b}})Wu, Huang, Jiang, Xie, Huang, and
  Zhao]{wu2025unfolding}
Weiqi Wu, Shen Huang, Yong Jiang, Pengjun Xie, Fei Huang, and Hai Zhao.
\newblock Unfolding the headline: Iterative self-questioning for news retrieval
  and timeline summarization.
\newblock \emph{arXiv preprint arXiv:2501.00888}, 2025{\natexlab{b}}.

\bibitem[Xia et~al.(2024)Xia, Deng, Dunn, and
  Zhang]{xia2024agentlessdemystifyingllmbasedsoftware}
Chunqiu~Steven Xia, Yinlin Deng, Soren Dunn, and Lingming Zhang.
\newblock Agentless: Demystifying llm-based software engineering agents, 2024.
\newblock URL \url{https://arxiv.org/abs/2407.01489}.

\bibitem[Yang et~al.(2024)Yang, Jimenez, Wettig, Lieret, Yao, Narasimhan, and
  Press]{yang2024sweagentagentcomputerinterfacesenable}
John Yang, Carlos~E Jimenez, Alexander Wettig, Kilian Lieret, Shunyu Yao,
  Karthik Narasimhan, and Ofir Press.
\newblock Swe-agent: Agent-computer interfaces enable automated software
  engineering.
\newblock \emph{Advances in Neural Information Processing Systems},
  37:\penalty0 50528--50652, 2024.

\bibitem[Yao et~al.(2023)Yao, Zhao, Yu, Du, Shafran, Narasimhan, and
  Cao]{yao2023react}
Shunyu Yao, Jeffrey Zhao, Dian Yu, Nan Du, Izhak Shafran, Karthik Narasimhan,
  and Yuan Cao.
\newblock React: Synergizing reasoning and acting in language models.
\newblock In \emph{International Conference on Learning Representations
  (ICLR)}, 2023.

\bibitem[Zhang et~al.(2024)Zhang, Ruan, Fan, and Roychoudhury]{autocoderover}
Yuntong Zhang, Haifeng Ruan, Zhiyu Fan, and Abhik Roychoudhury.
\newblock Autocoderover: Autonomous program improvement.
\newblock In \emph{Proceedings of the 33rd ACM SIGSOFT International Symposium
  on Software Testing and Analysis}, ISSTA 2024, pp.\  1592–1604, New York,
  NY, USA, 2024. Association for Computing Machinery.
\newblock ISBN 9798400706127.
\newblock \doi{10.1145/3650212.3680384}.
\newblock URL \url{https://doi.org/10.1145/3650212.3680384}.

\end{thebibliography}
